\documentclass[aps,prd,amssymb,eqsecnum,showpacs,nofootinbib,floatfix,twocolumn,a4paper]{revtex4-1}

\usepackage{amsmath}
\usepackage{graphicx}

\allowdisplaybreaks

\newcommand{\be}{\begin{equation}}
\newcommand{\ee}{\end{equation}}

\newcommand{\Pf}{{\mathrm{Pf}}}

\newcommand{\fii}{\phi_{(2)}}

\newcommand{\hTTiv}[2]{h^\mathrm{TT}_{(4)#1#2}}
\newcommand{\hTTivdot}[2]{\dot{h}^\mathrm{TT}_{(4)#1#2}}
\newcommand{\hTTivddot}[2]{\ddot{h}^\mathrm{TT}_{(4)#1#2}}
\newcommand{\ilhTTivddot}[2]{\Delta^{-1}[\ddot{h}^\mathrm{TT}_{(4)#1#2}]}

\newcommand{\md}{\mathrm{d}}

\newcommand{\pa}{\partial}

\newcommand{\pipi}{\mathbf{p}_1^2}
\newcommand{\pipip}{(\mathbf{p}_1^2)}
\newcommand{\pipii}{({\mathbf{p}}_1\cdot{\mathbf{p}}_2)}
\newcommand{\piipii}{{\bf p}_2^2}
\newcommand{\piipiip}{({\bf p}_2^2)}
\newcommand{\npi}{(\mathbf{n}_{12}\cdot\mathbf{p}_1)}
\newcommand{\npii}{({\bf n}_{12}\cdot{\bf p}_2)}

\def\npj{({\bf n}_{12}\cdot{\bf p}_2)}
\def\pjpj{{\bf p}_2^2}
\def\pjpjp{({\bf p}_2^2)}
\def\pipj{({\bf p}_1\cdot{\bf p}_2)}

\newcommand{\pp}{\mathbf{p}^2}
\newcommand{\ppn}{(\mathbf{p}^2)}
\newcommand{\np}{(\mathbf{n}\cdot\mathbf{p})}

\newcommand{\piTT}[2]{\pi^{#1#2}_\mathrm{TT}}

\newcommand{\pitiii}[2]{\widetilde{\pi}^{#1#2}_{(3)}}

\newcommand{\STTiv}[2]{S^\mathrm{TT}_{(4)#1#2}}

\def\htt#1#2{h^{\rm TT}_{#1#2}}
\def\httiv#1#2{h^{\rm TT}_{#1#2}}
\def\httivdot#1#2{\dot{h}^{\rm TT}_{#1#2}}

\begin{document}

\title{Nonlocal-in-time action for the fourth post-Newtonian conservative dynamics of~two-body systems}

\author{Thibault Damour}
\email{damour@ihes.fr}
\affiliation{Institut des Hautes Etudes Scientifiques, 35 route de Chartres, 91440 Bures-sur-Yvette, France}

\author{Piotr Jaranowski}
\email{pio@alpha.uwb.edu.pl}
\affiliation{Faculty of Physics,
University of Bia{\l}ystok,
Lipowa 41, 15--424 Bia{\l}ystok, Poland}

\author{Gerhard Sch\"afer}
\email{gos@tpi.uni-jena.de}
\affiliation{Theoretisch-Physikalisches Institut,
Friedrich-Schiller-Universit\"at,
Max-Wien-Pl.\ 1, 07743 Jena, Germany}


\begin{abstract}

We complete the analytical determination, at the 4th post-Newtonian (4PN) approximation,
of the conservative dynamics of gravitationally interacting two-point-mass systems.
This completion is obtained by resolving the infra-red ambiguity
which had blocked a previous 4PN calculation [P.~Jaranowski and G.~Sch\"afer, Phys.\ Rev.\ D \textbf{87}, 081503(R) (2013)]
by taking into account the 4PN breakdown of the usual near-zone expansion due to infinite-range tail-transported temporal correlations
found long ago [L.~Blanchet and T.~Damour, Phys.\ Rev.\ D \textbf{37}, 1410 (1988)].
This leads to a Poincar\'e-invariant 4PN-accurate effective action for two masses,
which mixes instantaneous interaction terms (described by a usual Hamiltonian) with a (time-symmetric) nonlocal-in-time interaction.

\end{abstract}


\maketitle

\section{Introduction}

The prospect of detecting, in the coming years, the gravitational wave signals emitted by coalescing binary systems of compact bodies
(neutron stars or black holes) provides a strong incentive for pushing the analytical theory of two-body systems to the highest possible accuracy.
Post-Newtonian (PN) theory is one of the key techniques for analytically describing the dynamics of binary systems.
Some time ago, the  conservative dynamics of binary systems has been obtained at the
3rd post-Newtonian (3PN) accuracy through a sequence of works
\cite{Jaranowski:1997ky,Jaranowski:1999,Damour:1999cr,Damour:2000kk,Jaranowski:2000,Blanchet:2000nv,Blanchet:2001,Damour:2000ni,deAndrade:2000gf}
that culminated in Ref.\ \cite{Damour:2001bu}
(see also \cite{Itoh:2003,Itoh:2004,Blanchet:2004,Foffa:2011} for later rederivations).
Recently, several works have obtained a partial knowledge of the conservative dynamics at the 4th post-Newtonian (4PN) accuracy 
\cite{Damour:2009sm,Blanchet:2010,Damourlogs,LeTiec:2012,Barausse:2011dq,Jaranowski:2012eb,Foffa:2012rn,Jaranowski:2013lca,Bini:2013zaa}
(see also \cite{Ledvinka:2008} for a closed-form expression valid to {\it all} PN-orders, at first order in Newton's gravitational constant).
We shall show here how to complete this line of work
by determining the full effective action describing the 4PN-accurate conservative two-body dynamics.

The stumbling block of Ref.\ \cite{Jaranowski:2013lca} was the appearance of irreducible infra-red (IR) divergences
in the calculation of the PN-expanded Hamiltonian $H_{\rm 4PN} (\mathbf{x}_1 , \mathbf{x}_2 , \mathbf{p}_1 , \mathbf{p}_2)$ of the binary system.
These IR divergences (separated as $H_{\rm 4PN}^{\rm inf} \equiv \int \md^3 x \, h_{\rm 4PN}^{\rm inf}$ in \cite{Jaranowski:2013lca})
will be further studied here and will be shown to be directly related to an old result of Blanchet and Damour \cite{Blanchet:1987wq}.
Reference \cite{Blanchet:1987wq} found that the usual PN scheme, based, in particular,
on a formal near-zone expansion of the flat-spacetime gravitational propagator, of the type
\begin{align}
\label{eq1.1}
&{\mathcal G}(t,\mathbf{x};t',\mathbf{x}') \equiv -4\pi \left( \Delta - \frac1{c^2} \, \partial_t^2 \right)^{-1}
\nonumber\\
&= -4\pi \left( \Delta^{-1} + \frac1{c^2}\Delta^{-2}\partial_t^2 + \frac1{c^4}\Delta^{-3}\partial_t^4 + \ldots \right) \delta (t-t')\,,
\end{align}
incurred a fundamental breakdown precisely at the 4PN level.
Indeed, at this level of accuracy it is crucial to take account of the fact that
the gravitational propagator ${\mathcal G}_g  (t, \mathbf{x} ; t' , \mathbf{x}')$ in the {\it curved spacetime} $g$ generated by the binary system contains,
even when both spatial positions $\mathbf{x}$ and $\mathbf{x}'$ are well within the usually defined near-zone
(i.e., when $\vert \mathbf{x} \vert , \vert \mathbf{x}' \vert \ll \lambdabar$
with $\lambdabar \equiv c \, \Omega^{-1}$ denoting the reduced wavelength associated to the orbital frequency $\Omega$),
a significant {\it tail} contribution whose support is not limited to lightlike intervals,
$\vert t-t' \vert \simeq \vert \mathbf{x} - \mathbf{x}' \vert / c$,
but extends to strongly time-nonlocal intervals $\vert t-t' \vert \gg \vert \mathbf{x} - \mathbf{x}' \vert / c$.
Reference \cite{Blanchet:1987wq} computed (for the case of the retarded propagator) the near-zone effect of these infinite-range tail-transported
temporal correlations, and we shall show below how the time-symmetric version of their result (related to the conservative part of the dynamics)
is precisely consistent with the IR divergences occurring when using (as was done in \cite{Jaranowski:2013lca})
the standard PN near-zone expansion Eq.~(\ref{eq1.1}).
This will allow us to remove these unphysical IR divergences and to replace them by their physical origin, a specific time-nonlocal interaction.

We employ the following notation:
$\mathbf{x}=\left(x^i\right)$ ($i=1,2,3$) denotes a point in the 3-dimensional
Euclidean space $\mathbb{R}^3$ endowed with a standard Euclidean metric
and a scalar product (denoted by a dot).
Letters $a$ and $b$ ($a,b=1,2$) are body labels,
so $\mathbf{x}_a\in\mathbb{R}^3$ denotes the position of the $a$th point mass.
We also define  ${\bf r}_a\equiv\mathbf{x}-\mathbf{x}_a$, $r_a \equiv |{\bf r}_a|$,
${\bf n}_a\equiv{\bf r}_a/r_a$; and for $a\ne b$, 
${\bf r}_{ab}\equiv\mathbf{x}_a-\mathbf{x}_b$,
$r_{ab} \equiv |{\bf r}_{ab}|$,
${\bf n}_{ab} \equiv {\bf r}_{ab}/r_{ab}$;
$|\cdot|$ stands here for the Euclidean length of a vector. 
The linear momentum vector of the $a$th body is denoted by $\mathbf{p}_a=\left(p_{ai}\right)$,
and $m_a$ denotes its mass parameter.
We abbreviate $\delta\left({\bf x}-{\bf x}_a\right)$ by $\delta_a$.
Extensive use has been made of the computer-algebra system \textsc{Mathematica}.

\section{Reduced (Fokker-type) action of a two-body system}

We are interested in the (reduced) action $S [x_1^{\mu},x_2^{\nu}]$
describing the conservative dynamics of an isolated, gravitationally interacting two-body system.
This Fokker-type action is formally obtained by eliminating the gravitational field $g_{\mu\nu}$,
conveying the time-symmetric (half-retarded-half-advanced) gravitational interaction,
in the total (gauge-fixed) action $S_{\rm tot} [x_a^{\mu} ; g_{\mu\nu}]$
describing the particles-plus-field system \cite{InfeldPlebanski,Ohta:1974pq,DS1985}.
When working in the harmonic gauge, the Fokker action can be written as an infinite series
$S_{\rm free} + S_{12} + \ldots\,$, where $S_{\rm free} = - \int m_1\,\md s_1 - \int m_2\,\md s_2$
(with $\md s_a = \sqrt{-\eta_{\mu\nu}\,\md x_a^{\mu}\,\md x_a^{\nu}}$) is the free action,
$S_{12}$ the one-graviton-exchange interaction \cite{Damour:1992we}
\begin{align}
\label{eq2.1}
S_{12} [x_1,x_2] &= 2G \iint \md s_1\md s_2 \, t_1^{\mu\nu}(s_1)
\nonumber\\
&\quad \times\,{\mathcal G}_{\mu\nu,\alpha\beta} (x_1 (s_1) - x_2 (s_2)) \, t_2^{\alpha\beta}(s_2) \,,
\end{align}
with linear source terms $t_a^{\mu\nu}(s_a) = m_a (\md x_a^{\mu}/\md s_a)(\md x_a^{\nu}/\md s_a)$,
gravitational propagator (in $D=4$ spacetime dimensions)
${\mathcal G}_{\mu\nu,\alpha\beta} = \left(\eta_{\mu\alpha} \, \eta_{\nu\beta} - \frac12 \, \eta_{\mu\nu} \, \eta_{\alpha\beta} \right) {\mathcal G}$,
with ${\mathcal G} (x,x') \equiv -4\pi \, \square_{\rm sym}^{-1} = \delta (\eta_{\mu\nu} (x^{\mu} - x'^{\mu})(x^{\nu} -x'^{\nu}))$,
and where the higher-order terms $+\ldots$ are given by more complicated Feynman-like integrals of the type (suppressing indices)
\begin{align}
\label{eq2.2}
S_{112} &\sim G^2  \iint\!\!\iint \md s_1 \, \md s'_1 \, \md s_2 \, \md^4 x \, t_1 (s_1)  \, t_1 (s'_1)  \, t_2 (s_2)
\nonumber \\
&\quad \times\,\partial \partial \, {\mathcal G} (x_1 - x) \, {\mathcal G} (x'_1 - x) \, {\mathcal G} (x - x_2) \, , 
\end{align}
where the concatenation of source terms, propagators, and vertices (here at the intermediate field point $x$)
is defined by the (gauge-fixed) Einstein-Hilbert action \cite{Damour:1995kt}.  The explicit form of the Poincar\'e-invariant
equations of motion at order $G^2$ has been obtained in Refs.\ \cite{Westpfahl:1979gu,Bel1981}.  
For the definition and computation of the PN-expanded version [using Eq.\ (\ref{eq1.1})] of the harmonic-gauge Fokker action
see Refs.\ \cite{Goldberger:2004jt,Gilmore:2008,Chu:2009,Foffa:2011}.

Previous works \cite{Ohta:1974kp,Schaefer:1986rd,Jaranowski:1997ky} have shown that a useful approach for computing the reduced gravitational action
is the canonical formalism of Arnowitt, Deser, and Misner (ADM) \cite{ADM62}.
There are less propagating degrees of freedom in this approach than in harmonic gauge.
Essentially $g_{00}$ and $g_{0i}$ have been eliminated, to leave only the spatial metric $g_{ij}$ and its canonically conjugated momentum $\pi^{ij}$.
The computation of the reduced two-body action (in spacetime dimension $D \equiv d+1$) within the ADM formalism goes through five steps.
Step (i) consists in fixing the gauge by requiring that $g_{ij}$ and $\pi^{ij}$ have the forms (ADMTT gauge)
\begin{subequations}
\label{eq2.3}
\begin{align}
g_{ij} &= A(\phi) \, \delta_{ij} + h_{ij}^{\rm TT} \, ,
\\[1ex]
\pi^{ij} &= \tilde\pi^{ij} (V^k) + \pi_{\rm TT}^{ij} \, , 
\end{align}
\end{subequations}
where
\begin{subequations}
\label{eq2.4}
\begin{align}
A(\phi) &\equiv \left( 1 + \frac{d-2}{4(d-1)} \, \phi \right)^{4/(d-2)} \, ,
\\
\tilde\pi^{ij}(V^k) &\equiv \partial_i \, V^j + \partial_j \, V^i - \frac2d \, \delta^{ij} \, \partial_k \, V^k \, , 
\end{align}
\end{subequations}
and where the TT pieces $h_{ij}^{\rm TT}$, $\pi_{\rm TT}^{ij}$ are transverse and traceless,
i.e., satisfy $\partial_j \, f_{ij}^{\rm TT} = 0 = \delta^{ij} \, f_{ij}^{\rm TT}$ with $f=h$ or $\pi$.

Step (ii) consists in solving with respect to $\phi$ and $V^i$ the Hamiltonian and momentum constraints,
i.e., (in units where $16\pi \, G_D = 1 = c$)
\begin{subequations}
\label{eq2.5}
\begin{align}
\sqrt{g}\,R &= \frac{1}{\sqrt g} \left(g_{ik} \, g_{j\ell} \, \pi^{ij} \, 
\pi^{k\ell} - \frac{1}{d-1} \ (g_{ij} \, \pi^{ij})^2 \right)
\nonumber\\[1ex]&\quad
+ \sum_a (m_a^2 + g_a^{ij} \, p_{ai} \, p_{aj})^{\frac{1}{2}} \, \delta_a,
\\[2ex]
-2\,D_j\,\pi^{ij} &= \sum_a g_a^{ij} \, p_{aj} \, \delta_a. 
\end{align}
\end{subequations}
Here, the usual geometrical quantities (spatial scalar curvature $R$, spatial covariant derivative $D_j$, \ldots)
refer to a $d$-dimensional space, and $g_a^{ij}$ denotes $g^{ij} (\mathbf{x}_a)$.
We dimensionally continue $d$ in the complex plane before letting $d$ tend back to 3 at the end of the calculation.

The constraints (\ref{eq2.5}) yield an elliptic system for $\phi$ and $V^i$ which has the structure
\begin{subequations}
\label{eq2.6}
\begin{align}
&\Delta \phi = -\sum_a m_a (1+\ldots) \, \delta_a + \ldots \,,
\\
&\Delta V^i + \left( 1 - \frac2d \right) \partial_{ij} \, V^j = -\frac12 \sum_a (p_{ai} + \ldots ) \, \delta_a + \ldots \,.
\end{align}
\end{subequations}
One can perturbatively solve this system in powers of $m_a , p_{ai}$ and of $h^{\rm TT}$ and $\pi_{\rm TT}$ (that enter the ellipsis).

Step (iii) then consists in computing the Hamiltonian of the total particles-plus-field system
\be
\label{eq2.7}
H_\text{tot}\big[\mathbf{x}_a,\mathbf{p}_a,{\htt ij},{\piTT ij}\big]
= -\int \md^dx\,\Delta\phi\big[\mathbf{x}_a,\mathbf{p}_a,{\htt ij},{\piTT ij}\big].
\ee
Two more steps are then needed to derive the reduced action for the particles.
One must Legendre transform the above Hamiltonian with respect to the field variables
to get the ``Routhian'' \cite{Jaranowski:1997ky,Jaranowski:2012eb,Jaranowski:2013lca},
\be
\label{eq2.8}
R\big[\mathbf{x}_a,\mathbf{p}_a,{\httiv ij},{\httivdot ij}\big]
\equiv H_\text{tot}  - \int\md^dx\,{\piTT ij}{\httivdot ij}.
\ee
Finally, the reduced (Fokker-type) action for the particle system (in Hamiltonian form) is
$S = \sum_a\int \mathbf{p}_a\,\md\mathbf{x}_a - \int\md t \, H[\mathbf{x}_a , \mathbf{p}_a]$,
where the particle Hamiltonian $H[\mathbf{x}_a,\mathbf{p}_a]$ is formally obtained
by ``integrating out'' the field variables $h_{ij}^{\rm TT} , \dot h_{ij}^{\rm TT}$,
i.e., by replacing them by their solutions as a functional of the particle variables
\be
\label{eq2.9}
H[\mathbf{x}_a,\mathbf{p}_a] = R\big[ {\bf x}_a,{\bf p}_{a},
{\httiv ij}(\mathbf{x}_a,\mathbf{p}_a), {\httivdot ij}(\mathbf{x}_a,\mathbf{p}_a) \big] \, .
\ee 
We shall discuss below the subtleties linked to this formal elimination of the field variables
(beyond the well-understood elimination of higher-order time derivatives of $\mathbf{x}_a$ and $\mathbf{p}_a$
through the use of lower-order equations of motion \cite{Schafer84,DS1985,DS1991}).

\section{IR ambiguity in the near-zone expansion of the 4PN reduced action}

Equations (\ref{eq2.1}) and (\ref{eq2.2}) for the general structure (in harmonic coordinates)
of the reduced two-body action $S[x_1^{\mu} (s_1), x_2^{\nu} (s_2)]$ clearly show that this action is, a priori, nonlocal-in-time,
i.e., is a functional of the two world lines $x_1^{\mu} (s_1)$, $x_2^{\nu} (s_2)$
which involves arbitrarily large proper-time separations $\vert s_1-s_2 \vert$.
In the ADM gauge, this nonlocality is less severe because the PN most prominent field degrees of freedom ($\phi$ and $V^i$)
have instantaneous propagators [see Eq.~\eqref{eq2.6}].
However, the time nonlocality arises when one integrates out $h_{ij}^{\rm TT}$ and $\dot h_{ij}^{\rm TT}$,
because these field variables propagate at the velocity of light.
At low PN orders (up to 3PN included) it is possible (by using many integrations by parts) to express the reduced Hamiltonian
entirely in terms of $\phi_{(2)}$, $\phi_{(4)}$, $V^i_{(3)}$, $h_{(4)ij}^{\rm TT}$, and $\dot h_{(4)ij}^{\rm TT}$ \cite{Jaranowski:1997ky,Damour:2001bu}.
[Here, the  numbers within parentheses denote the formal order in the inverse velocity of light, e.g., $\phi_{(2)} \sim G m / (c^2 \, r^{d-2})$.]
The elimination of $h_{(4) ij}^{\rm TT}$ and $\dot h_{(4)ij}^{\rm TT}$ can then be done by means of instantaneous propagators, using
\begin{align}
\label{eq3.1}
\Delta h_{(4) ij}^{\rm TT} &= \STTiv{ij} \,,
\end{align}
with the source term
\be
\label{eq3.1a}
S_{(4)ij} \equiv -\sum_a \frac{p_{ai} \, p_{aj}}{m_a} \, \delta_a - \frac{d-2}{2(d-1)} \, \phi_{(2),i} \, \phi_{(2),j} \,.
\ee
After this the 3PN-accurate Hamiltonian can be obtained (by computing an IR-convergent spatial integral)
as a local-in-time function of $\mathbf{x}_a$ and $\mathbf{p}_a$.

The situation changes at the 4PN level.
At this level there appear (in any gauge) irreducible IR divergences.
However, we could isolate the IR divergences in a few contributions.
By using many integrations by parts (both in space and in the time domain),
we could decompose the 4PN-level integrand for the Routhian, Eq.~(\ref{eq2.8}), into two parts
(details of the computation will be published elsewhere \cite{JS2014}), say
\be
\label{eq3.2}
{\mathit{r}}_{\mathrm{4PN}} = {\mathit{r}}_{\mathrm{4PN}}^1 + {\mathit{r}}_{\mathrm{4PN}}^2.
\ee
The part ${\mathit{r}}_{\mathrm{4PN}}^1$ (which contains most of the contributions)  is made of terms that are either 
IR convergent or whose IR behavior can be unambiguously regularized to zero. 
The part ${\mathit{r}}_{\mathrm{4PN}}^2$  collects the few terms that  contain ambiguous  IR divergences 
(generating logarithms)  at $r \equiv \vert\mathbf{x}\vert \to \infty$. 
It explicitly reads
\begin{align}
\label{r2}
{\mathit{r}}_{\mathrm{4PN}}^2 &=
\frac{1}{2(d-1)}\fii{\hTTiv ij}{\hTTivddot ij}
\nonumber\\[1ex] &\quad
-\frac{1}{4}{\hTTivdot ij} \Delta^{-1}[\dddot{h}_{(4)ij}^\textrm{TT}]
+ \bigg(\frac{1}{2(d-1)}\fii(\Delta{\hTTiv ij})
\nonumber\\[1ex] &\quad
+ \frac{d-2}{d-1}\frac{\pa}{\pa t}\big(\fii{\pitiii ij}\big) \bigg) {\ilhTTivddot ij}.
\end{align}
From the technical point of view, the crucial feature of Eq.~(\ref{eq3.2})
is that all its  terms (involving various inverse Laplacians)
can be explicitly computed, and that the corresponding contributions to the Hamiltonian (obtained by integrating over space)
can also be fully evaluated (in $d=3$ dimensions) by using techniques
developed in Refs.\ \cite{Jaranowski:1997a,Jaranowski:1997b,Jaranowski:1997ky}.

Many of the integrals (both in  ${\mathit{r}}_{\mathrm{4PN}}^1$ and in $ {\mathit{r}}_{\mathrm{4PN}}^2$) contain ultra-violet (UV) divergences,
i.e., divergences near the particles, as $r_1 = \vert \mathbf{x} - \mathbf{x}_1 \vert \to 0$ or $r_2 = \vert\mathbf{x}-\mathbf{x}_2\vert \to 0$.
All the UV divergences are conveniently regularized by using dimensional regularization,
in the way described in Refs.~\cite{Damour:2001bu,Jaranowski:2013lca}:
i.e., by computing the (locally generated) difference between the dimensionally regularized integral
and its Riesz-implemented Hadamard-regularized version.
The important result (reported in \cite{Jaranowski:2013lca}) is that all the UV divergences [i.e., all the poles in $1/(d-3)$]
can be removed from the Hamiltonian by adding a total time derivative.

From the conceptual point of view, Eq.~(\ref{eq3.2}) is partly unsatisfactory because,
for eliminating $h_{ij}^{\rm TT}$ and $\dot h_{ij}^{\rm TT}$,
it used the (time-symmetric) PN expansion (\ref{eq1.1}) of the propagator $\square^{-1}$ entering the TT propagator for $h_{ij}^{\rm TT}$.
In particular, it is the $\mathcal{O}\left(c^{-2}\right)$ correction term in Eq.~(\ref{eq1.1})
which is responsible for the appearance of all the terms in Eq.\ \eqref{r2},
which have the following structure
\be
\label{eq3.3}
f_{ij} (\mathbf{x}) \, \Delta^{-1} \left(\ddot h_{(4)ij}^{\rm TT} \right)
= f_{ij}(\mathbf{x}) \, \Delta^{-2} \, \partial_t^2 \, {\STTiv ij}
\ee
[the second term in Eq.\ \eqref{r2} has also this structure
modulo an integration by parts with respect to time].
Before discussing in detail the physical meaning of these IR divergences,
let us study the ambiguities arising when formally regulating them.
We used several ways to regulate these IR divergences
(while maintaining the possibility to explicitly compute the Hamiltonian).
In all cases, one needs to introduce a new length scale, say $s$.
For instance, in all the problematic terms of the type (\ref{eq3.3})
[including the factor $\ddot h_{(4)ij}^{\rm TT}$ in Eq.~(\ref{r2}),
rewritten as $\Delta (\Delta^{-1} \, \ddot h_{(4)ij}^{\rm TT})$]
one can make the replacement
\be
\label{eq3.5}
\Delta^{-1} \left[ \ddot h_{(4)ij}^{\rm TT}\right]
\to \Delta^{-1} \left[ \left( \frac rs \right)^B \, \ddot h_{(4)ij}^{\rm TT}\right]^\textrm{TT}
\ee
and then take the finite part of the pole occurring at $B=0$ in 3 dimensions
(and displaced at $B=2(d-3)$ in $d$ dimensions; see Sec.~VIII in \cite{Blanchet:2005tk}).
Alternatively one can multiply, before integrating it over space,
the full integrand by a factor $\left( \frac{r_1}s \right)^{\!\alpha} \left( \frac{r_2}s \right)^{\!\beta}$
and take the finite part of the IR pole occurring at $\alpha + \beta = 2(d-3)$.
Both methods conveniently allow one to detect, and subtract,
the {\it logarithmic} IR divergence linked to a decay of (parts of) the integrand $\propto r^{-3-3(d-3)}$ as $r \to \infty$.
We have checked that both methods yield the same result modulo (a time derivative and) a change in the constant $C$ introduced below.
We shall denote the result of the specific IR-regularization  (\ref{eq3.5})
of the (separately UV-regularized, as explained above) reduced PN-expanded Hamiltonian as $H_{\rm 4PN}^\textrm{near-zone\,($s$)}$.
By explicitly calculating this near-zone-related 4PN-accurate Hamiltonian
[which used the formal, IR-delicate, near-zone expansion (\ref{eq1.1}),
regulated by an IR scale $s$, say as in Eq.~(\ref{eq3.5})], we found that it has the structure
\begin{align}
\label{eq3.6}
H_{\rm 4PN}^\textrm{near-zone\,($s$)}[\mathbf{x}_a,\mathbf{p}_a] &= H_{\rm 4PN}^{{\rm loc} \, 0}[\mathbf{x}_a,\mathbf{p}_a]
\nonumber \\
&\quad + F[\mathbf{x}_a,\mathbf{p}_a] \left( \ln \, \frac{r_{12}}s + C \right)
\nonumber \\
&\quad + \frac{\md}{\md t}\,G[\mathbf{x}_a,\mathbf{p}_a],
\end{align}
where the coefficient of the IR-dependent logarithm is equal
(after separating some total time derivative, incorporated in the last term) to
\be
\label{eq3.7}
F[\mathbf{x}_a,\mathbf{p}_a] = \frac25 \ \frac{G^2 M}{c^8} \big(I_{ij}^{(3)}\big)^2 \,.
\ee
Here $M:= m_1 + m_2$, the superscript $(3)$ denotes a third time derivative,
and $I_{ij}$ denotes the (Newtonian) quadrupole moment of the binary system
\be
I_{ij} := \underset{a}{\sum} \, m_a \left(x_a^i \, x_a^j - \frac13 \, \delta^{ij} \, \mathbf{x}_a^2 \right).
\ee
Here, and below [as well as in Eq.~(\ref{eq2.9}) above], we use brackets, rather than parentheses,
around the dynamical arguments $\mathbf{x}_a$, $\mathbf{p}_a$ to signal that the considered quantity
might depend not only on the instantaneous values of $\mathbf{x}_a$ and $\mathbf{p}_a$,
but also on several of their time derivatives
[and even, in the case of Eq.~(\ref{eq2.9}), on the full time evolution of the dynamical variables].

We have added to the logarithm in Eq.~(\ref{eq3.6}) an arbitrary constant $C$ to remind us that the IR-regularization scale is arbitrary.
(Replacing $s$ by $s'=e^{-\lambda}\,s$ is equivalent to replacing $C$ by $C'=C+\lambda$.)

\section{Additional tail contribution to the 4PN reduced action}

The result (\ref{eq3.6}) given by the usual PN approximation scheme can only be an incomplete representation of the two-body conservative dynamics
because it depends (for a given choice of the scale $s$) on the arbitrary constant $C$.
This incompleteness of a (near-zone limited) 4PN-level calculation is in precise accord
with an old result of Blanchet and Damour \cite{Blanchet:1987wq} (see \cite{Foffa:2013} for a recent rederivation).
Indeed, Ref.~\cite{Blanchet:1987wq} found that, precisely at the 4PN level,
there occurred a fundamental breakdown of one of the basic tenets of the usual post-Newtonian approximation scheme.
At the 4PN level, it becomes impossible (in any gauge) to express the near-zone metric
(and therefore, also, the two-body equations of motion)
as a functional of the  instantaneous state of the material source.
Because of correlations transported over arbitrarily large time differences by tail effects (viewed in the near-zone),
the equations of motions at time $t$ depend on the state of the system at all times $t' < t$ (when considering retarded interactions).
[Such long-range correlations are already a priori contained in the Fokker-action contributions such as Eq.~(\ref{eq2.2}).
However, the work of \cite{Blanchet:1987wq} shows that this becomes physically important only at the 4PN level.]
Reference~\cite{Blanchet:1987wq} used a technique of matching between the near-zone $r \ll \lambdabar$ (where PN expansions should be adequate)
and the exterior zone $r \gg r_{12}$ (where multipolar post-Minkowskian (MPM) expansions \cite{Blanchet:1985sp} are adequate)
to compute the near-zone effect of tail-transported correlations.
Their result depends on an arbitrary length scale that they denoted $r_1 \equiv cP$.
As we have allowed for an arbitrary additional constant $C$ in Eq.~(\ref{eq3.6}),
we can, and will, identify the length scale $r_1 \equiv cP$ used in \cite{Blanchet:1987wq} with the length scale $s$ used in the previous section.
We note that the length scale $r_1 \equiv cP \equiv s$ was introduced in the MPM formalism \cite{Blanchet:1985sp}
in a form very similar to Eq.~(\ref{eq3.5}) above.
Though this length scale is arbitrary it can play the role (both here, and in Refs.~\cite{Blanchet:1985sp,Blanchet:1987wq})
of an intermediate scale between the scale of the system $r_{12}$ and the wavelength $\lambdabar = c/\Omega$.
Indeed, if $r_{12} \ll s \ll \lambdabar$ both the PN expansion and the MPM one should be valid
(and can be matched to each other) at distances $r \sim s$.

Independently of this interpretation, the main result of \cite{Blanchet:1987wq} was their Eq.~(6.33),
saying that 4PN-level inner metric is the sum of a PN-like instantaneous functional of the source variables $h_{\mu\nu}^{\rm inst}$
(involving the logarithm of $s$), and of a specific nonlocal-in-time ``tail'' contribution equal (in a suitable gauge) to
\begin{subequations}
\label{eqs4.1-4.3}
\begin{align}
\label{eq4.1}
h_\textrm{00\,4PN}^{{\rm tail} \, (s)} (t,\mathbf{x}) &= h_\textrm{00\,4PN}^{{\rm tail \, sym} \, (s)} + h_\textrm{00\,4PN}^{\rm rad \, reac} \, ,
\\[1ex]
\label{eq4.2}
h_\textrm{00\,4PN}^{{\rm tail \, sym} \, (s)} &= -\frac45 \, \frac{G^2 M}{c^{10}} \, x^i \, x^j
\nonumber\\
\times\,\Pf_{2s/c} & \left[ \int_0^{+\infty} \frac{\md v}v \left( I_{ij}^{(6)} (t-v) + I_{ij}^{(6)} (t+v) \right)\right] \,,
\\
\label{eq4.3}
h_\textrm{00\,4PN}^{\rm rad \, reac} &= -\frac45 \, \frac{G^2 M}{c^{10}} \, x^i \, x^j
\nonumber\\
\times\int_0^{+\infty} & \frac{\md v}v \left( I_{ij}^{(6)} (t-v) - I_{ij}^{(6)} (t+v) \right) \,.
\end{align}
\end{subequations}
Here, we have integrated by parts and decomposed the retarded-propagator result of \cite{Blanchet:1987wq}
in its time-symmetric (conservative) and time-antisymmetric (radiation-reaction \cite{Blanchet:1993ng}) parts.
The symbol $\Pf_T$ denotes a Hadamard partie finie with time scale $T$ (with $T:=2s/c\equiv 2P$)
\begin{align}
\Pf_T \int_0^{+\infty} \frac{\md v}v \, g(v) &:= \int_0^T \frac{\md v}v \, \big(g(v)-g(0)\big)
\nonumber\\[1ex]
&\quad + \int_T^{+\infty} \frac{\md v}v \, g(v) \, .
\end{align}
Equivalently, $\Pf_T$ can be defined as the finite part in the Laurent expansion around $B=0$ ($\mathrm{FP}_B$)
of the analytic continuation in the complex parameter $B$ of the integral obtained by multiplying
the integrand $v^{-1}g(v)$ by a factor $(|v|/T)^B$.
\big[This second definition can be more convenient when working, as we shall do below,
with two-sided integrals $\mathrm{FP}_B\int_{-\infty}^{+\infty} \md v (|v|/T)^B (\ldots)$,
or with double integrals $\mathrm{FP}_B \int\int\md t\,\md t' (|t-t'|/T)^B (\ldots)$\big].

The contribution to the equations of motion of the two-body system
following from the time-symmetric 4PN tail metric derives from the action
\be
\label{eq4.4}
\frac12 \times \frac12 \int \md t
\sum_a m_a c^2 \, h_\textrm{00\,4PN}^{{\rm tail \, sym}}(t,\mathbf{x}_a),
\ee
where the extra factor $\frac12$ (beyond the usual $U = \frac12 \, h_{00}$ one)
comes from the symmetric bilinear functional dependence of $S^\textrm{tail sym}$ on $I_{ij}(t)$.
Inserting (\ref{eq4.2}) into (\ref{eq4.4}) and operating three times by parts yields
\begin{align}
\label{eq4.5}
S_{\rm 4PN}^\textrm{tail sym\,($s$)} &= +\frac15 \, \frac{G^2 M}{c^8}
\nonumber \\
&\quad \times\,\Pf_{2s/c} \iint \frac{\md t\,\md t'}{\vert t-t' \vert} \, I_{ij}^{(3)} (t) \, I_{ij}^{(3)} (t') \, .
\end{align}
(Reference \cite{Foffa:2013} considered a related, but different,
action depending, \`a la Schwinger-Keldysh, on a ``doubled'' quadrupole moment.)
The action (\ref{eq4.5}) formally corresponds to a nonlocal Hamiltonian
$\left( S_{\rm 4PN}^\textrm{tail sym\,($s$)} = - \int\md t\,H_{\rm 4PN}^\textrm{tail sym\,($s$)}(t)\right)$ equal to
\begin{align}
\label{eq4.6}
H_{\rm 4PN}^\textrm{tail sym\,($s$)}(t) &= -\frac15 \, \frac{G^2 M}{c^8} \,I_{ij}^{(3)} (t)
\nonumber \\
&\quad \times\,\Pf_{2s/c} \int_{-\infty}^{+\infty} \frac{\md v}{\vert v \vert} \, I_{ij}^{(3)} (t+v) \,.
\end{align}

Combining the result (\ref{eq3.6}) of the previous section
(corresponding to the effect of $h_{\mu\nu}^{\rm inst}$, left undetermined in \cite{Blanchet:1987wq}),
with the additional nonlocal term (\ref{eq4.6}),
we conclude that the two-body action describing the conservative 4PN dynamics
must correspond to the nonlocal Hamiltonian
\be
\label{eq4.7}
H_{\rm 4PN}^{\rm tot} = H_{\rm 4PN}^\textrm{near-zone\,($s$)} + H_{\rm 4PN}^\textrm{tail sym\,($s$)}\, .
\ee

A first indication of the correctness of this result is that the dependence on the arbitrary scale $s$
cancels between the two contributions on the right-hand side of (\ref{eq4.7}).
Indeed, the $s$-dependence of the tail contribution (\ref{eq4.6}) is easily seen to be
\be
\label{eq4.8}
H_{\rm 4PN}^{{\rm tail \, sym} \, (s)} = +\frac25 \, \frac{G^2 M}{c^8} \left(I_{ij}^{(3)} (t)\right)^2 \ln (2s/c) + \ldots 
\ee
to be compared with Eqs.\ (\ref{eq3.6}) and (\ref{eq3.7}).
Note that the scale $s$ is a UV cutoff (small $v$) in 
$H_{\rm 4PN}^\textrm{tail sym\,($s$)}$, and an IR one (large $r$) in $H_{\rm 4PN}^\textrm{near-zone\,($s$)}$.
This confirms the usefulness of thinking of $s$ as being an intermediate scale between 
the size of the system $r_{12}$ and the wavelength $\lambdabar = c/\Omega$
(similar to the introduction of an intermediate scale
when decomposing the calculation of the Lamb-shift in two complementary parts). 

On the other hand, the dependence of (\ref{eq4.7}), via (\ref{eq4.8}), on $\ln c$ is meaningful
and agrees with the logarithms arising at 4PN in the two-body dynamics
\cite{Damour:2009sm,Blanchet:2010,Damourlogs,LeTiec:2012,Barausse:2011dq,Jaranowski:2012eb,Jaranowski:2013lca,Bini:2013zaa}.

\section{Completion of the determination of the 4PN reduced action}

The 4PN contribution to the reduced action,
\begin{align}
\label{eq5.1}
S_{\rm 4PN} &= \frac15 \, \frac{G^2 M}{c^8} \, \Pf_{2s/c}
\iint \frac{\md t \, \md t'}{\vert t-t' \vert} \, I_{ij}^{(3)}(t) \, I_{ij}^{(3)}(t')
\nonumber\\[1ex]&\quad
- \int \md t \, H_{\rm 4PN}^{\textrm{near-zone\,($s$)}},
\end{align}
still contains an unknown constant $C$, entering Eq.~(\ref{eq3.6}).
To determine it analytically we need a calculation which fully takes into account the transition between the near zone and the wave zone,
without losing any information in the process. Such a calculation was recently performed in Ref.~\cite{Bini:2013zaa},
in the particular case of the dynamics of circular orbits.
Before using this result to complete the analytic determination of the 4PN two-body action,
let us report on a very satisfactory feature of the Hamiltonian (\ref{eq4.7}) [and the action (\ref{eq5.1})].
We have explicitly computed, in an arbitrary (nonmass-centered) frame,
the quite complicated  main contribution $H_{\rm 4PN}^{\rm loc\,0} [\mathbf{x}_a,\mathbf{p}_a]$
to the Hamiltonian in Eq.~(\ref{eq3.6}) (see Appendix A).
It has a polynomial structure in $\mathbf{p}_a$, $1/r_{12}$ and $\mathbf{n}_{12}$ of the type
\be
\label{eq5.2}
H_{\rm 4PN}^{\rm loc\,0} \sim p^{10} + \frac{p^8}{r_{12}} + \frac{p^6}{r_{12}^2}
+ \frac{p^4}{r_{12}^3} + \frac{p^2}{r_{12}^4} + \frac1{r_{12}^5} \,.
\ee
We have then proven that the 4PN-accurate dynamics, defined by
$H^{\mathrm{loc}\,0}_\mathrm{\leq 4PN} = Mc^2 + H_\textrm{N} + H_{\rm 1PN} + H_{\rm 2PN} + H_{\rm 3PN} + H_{\rm 4PN}^{\rm loc\,0}$,
was Poincar\'e-invariant in the usual sense \cite{Damour:2000kk} of admitting ten conserved quantities
[and notably the crucial boost generator 
$K^i(\mathbf{x}_a,\mathbf{p}_a,t) = G^i(\mathbf{x}_a,\mathbf{p}_a) - t\,P^i(\mathbf{x}_a,\mathbf{p}_a)$]
whose brackets realize the full (PN-expanded) Poincar\'e algebra.
In addition, as $r_{12}$, $F$ [Eq.\ \eqref{eq3.7}],
and also $I_{ij}^{(3)}(t)\,I_{ij}^{(3)}(t')$ are Galileo invariant,
both the logarithmic local contribution $F\biglb(\ln(r_{12}/s) +C\bigrb)$ in $H_{\rm 4PN}^{\textrm{near-zone\,($s$)}}$
and, formally, the nonlocal tail contribution \eqref{eq4.5} to the action
are consistent with Poincar\'e invariance, independently of the value of the constant $C$. 

In view of this result, we shall write down in the main text only the simpler center-of-mass expression of $H_{\rm 4PN}$
[see Appendix A for the expression of the general-frame Hamiltonian $H(\mathbf{x}_a,{\mathbf p}_a)$,
as well as the center-of-energy vector $G^i({\bf x}_a,{\bf p}_a)$].
Before doing this, let us indicate how to determine the value of $C$.
The simplest way to determine it is to compare the (gauge-invariant) functional link $E(j;\nu)$
between the 4PN-accurate binding energy $H-Mc^2$
and the (reduced, dimensionless) angular momentum $j := c \, J/(G \, m_1 \, m_2)$ along circular orbits, predicted by our $H_{\rm 4PN} (C)$,
to the corresponding result derived from the effective one-body formalism \cite{Buonanno:1998gg,DJS2000},
when using the recently determined 4PN-accurate radial potential $A_\textrm{4PN}(u)$ \cite{Bini:2013zaa}.
(Note that our determination of the value of $C$ does not rely on the various resummations entering the effective one-body formalism,
but only on its PN-expanded Hamiltonian content.)

Using the notation $j:=c \, J / (G \, m_1 \, m_2)$, $\mu := m_1 \, m_2 / (m_1 + m_2)$,
and $\nu := \mu/M = m_1 \, m_2 / (m_1 + m_2)^2$,
the 4PN-accurate effective one-body radial potential \cite{Bini:2013zaa} yields
\begin{widetext}
\begin{align}
\label{eq5.3}
E_\textrm{$\le$4PN}(j;\nu) &= - \frac12\,\mu c^2 \frac1{j^2} \Bigg( 1+\frac14 \, (9+\nu) \, \frac1{j^2} 
+ \frac{1}{8} (81- 7\nu + \nu^2) \frac{1}{j^4}
+ \bigglb( \frac{3861}{64}+\bigg(\frac{41\pi^2}{32}-\frac{8833}{192}\bigg) \nu -\frac{5\nu^2}{32}+\frac{5\nu^3}{64} \biggrb) \frac{1}{j^6}
\nonumber\\[1ex]&\quad
+ \bigglb(\frac{53703}{128} + \bigg(\frac{6581\pi^2}{512}-\frac{989911}{1920}-\frac{64}{5}\bigg(2\gamma_\mathrm{E}+\ln\frac{16}{j^2}\bigg)\bigg)\nu
+ \bigg(\frac{8875}{384}-\frac{41\pi^2}{64}\bigg)\nu^2-\frac{3\nu^3}{64}+\frac{7\nu^4}{128}\biggrb) \frac{1}{j^8} \Bigg),
\end{align}
where $\gamma_\mathrm{E}=0.577\ldots$ denotes Euler's constant.
Let us note that the above 4PN-accurate expansion of $E(j;\nu)$ contains 
(when using the relation  $\md E = \Omega\,\md J$) the same information
as the sometimes used PN expansion $E(x;\nu)$ of the binding energy
as a function of the dimensionless frequency parameter $x:=(G M \Omega / c^3)^{2/3}$.
Explicitly, we have the 4PN-accurate links
\begin{subequations}
\begin{align}
x(j;\nu) &= \frac1{j^2} \Bigglb( 1 + \bigg(3 + \frac{\nu}{3}\bigg)\frac1{j^2}
+ \bigg(18-\frac{9 \nu }{4}+\frac{2 \nu ^2}{9}\bigg)\frac1{j^4}
+ \bigglb(135+\left(\frac{41 \pi ^2}{12}-\frac{8779}{72}\right) \nu -\frac{\nu ^2}{3}+\frac{14 \nu ^3}{81}\biggrb)\frac1{j^6}
\nonumber \\[1ex]&\quad
+ \Bigg( 1134 + \bigglb(\frac{28969\pi^2}{768}-\frac{4449821}{2880}-\frac{128}{3} \bigg(2\gamma_\mathrm{E}+\ln\frac{16}{j^2}\bigg)\biggrb)\nu
+ \left(\frac{20399}{216}-\frac{779\pi^2}{288}\right)\nu^2 - \frac{5 \nu ^3}{54} + \frac{35 \nu^4}{243} \Bigg)\frac1{j^8}
\Biggrb),
\\[1ex]
\frac1{j(x;\nu)^2} &= x \Bigg( 1 -\bigg(3+\frac{\nu }{3}\bigg) x
+ \frac{25}{4} \nu\,x^2
+ \bigg(\frac{5269}{72}-\frac{41 \pi ^2}{12}-\frac{61\nu}{12}+\frac{\nu ^2}{81}\bigg) \nu\,x^3
\nonumber \\[1ex]
&\quad + \bigglb(\frac{18263\pi^2}{768}-\frac{1294339}{2880}+\frac{128}{3}\big(2\gamma_\mathrm{E}+\ln(16 x)\big)
+\left(\frac{2747\pi^2}{288}-\frac{90985}{432}\right) \nu +\frac{181\nu^2}{108}+\frac{\nu^3}{243}\biggrb) \nu\,x^4 \Bigg).
\end{align}
\end{subequations}
from which follows
\begin{align}
E_\textrm{$\le$4PN}(x;\nu) = -\frac{\mu c^2 x}{2} \Bigg(
1 &-  \bigg(\frac{3}{4} + \frac{\nu}{12} \bigg) x
+ \bigg(-\frac{27}{8} + \frac{19\nu}{8} - \frac{\nu^2}{24}\bigg) x^2
\nonumber\\[1ex]&
+ \bigg(-\frac{675}{64} + \left(\frac{34445}{576}-\frac{205\pi^2}{96}\right)\nu -\frac{155\nu^2}{96} - \frac{35\nu^3}{5184} \bigg) x^3
\nonumber\\
&+ \bigg( -\frac{3969}{128}
+ \bigg(\frac{9037 \pi ^2}{1536}-\frac{123671}{5760}+\frac{448}{15}\big(2\gamma_\mathrm{E}+\ln(16 x)\big)\bigg)\nu
\nonumber\\
&\qquad + \left(\frac{3157\pi^2}{576} -\frac{498449}{3456}\right)\nu^2
+ \frac{301\nu^3}{1728} + \frac{77\nu^4}{31104} \bigg) x^4 \Bigg).
\end{align}
\end{widetext}
It is straightforward to derive the $E(j;\nu)$ link following from our nonlocal Hamiltonian (\ref{eq4.7}).
It involves the evaluation of the nonlocal piece (\ref{eq4.6}) along circular motion (without any differentiation).
We proceed as follows.
Combining the $s$-dependent piece (\ref{eq4.6}) with the $F\ln(r_{12}/s)$ piece in (\ref{eq3.6})
(which has the effect of replacing the scale $s$ by the scale $r_{12}$) yields the integral
\be
\label{eq5.4}
H_{\rm 4PN}^{{\rm tail \, sym}\,(r_{12})}
= -\frac15 \, \frac{G^2 M}{c^8} \, \Pf_{2r_{12}/c} \int_{-\infty}^{+\infty} \frac{\md v}{\vert v \vert} \, f(v),
\ee
where $f(v) := I_{ij}^{(3)} (t) \, I_{ij}^{(3)} (t+v)$.
Along a circular motion one has $f(v) = f(0) \cos(2\Omega v)$, where
\be
\label{eq5.7}
f(0) = \big(I_{ij}^{(3)}(t)\big)^2 = 32 \, (\mu \, \Omega^3 \, r_{12}^2 )^2 \, .
\ee
Using the result (for $\omega>0$)
\be
\label{eq5.5}
\Pf_{T} \int_{0}^{+\infty} \frac{\md v}{v} \cos \, (\omega v) = - \big(\gamma_\mathrm{E} + \ln(\omega\,T)\big) \,,
\ee
Eq.\ \eqref{eq5.4} can be written as
\be
\label{eq5.6}
H_{\rm 4PN}^{{\rm tail \, sym} \, (r_{12})}
= + \frac25 \, \frac{G^2 M}{c^8} \, f(0) \left(\gamma_\mathrm{E} + \ln \left( \frac{4\,\Omega\,r_{12}}{c} \right)\right).
\ee

Note how the effective replacement of the arbitrary, intermediate scale $s$ by $r_{12}$ in the tail contribution
has generated the combination $\gamma_\mathrm{E}+\ln4$ accompanying $\ln(\Omega\,r_{12}/c)=-\ln j$ in Eq.~(\ref{eq5.3}).
In addition, we found that the $\pi^2$ contribution associated to these terms is already contained in $H_{\rm 4PN}^{\rm loc \, 0}$.
Finally, when precisely defining the IR-regularized piece $H_\textrm{4PN}^{\textrm{loc\,0}}$ by the procedure (\ref{eq3.5}),
we find that the Hamiltonian (\ref{eq3.6}) (where the total derivative term does not contribute) yields a circular link $E(j;\nu)$
in full agreement with Eq.~(\ref{eq5.3}) if the constant $C$ in (\ref{eq3.6}) is equal to the rational number
\be
\label{eq5.8}
C = -\frac{1681}{1536}\,.
\ee
This result completes the determination of the 4PN conservative dynamics.

Let us summarize our results by writing the total 4PN Hamiltonian in its center-of-mass form
(in terms of the reduced variables $\mathbf{r} := \mathbf{x}_{12}/(GM)$, $\mathbf{p} := \mathbf{p}_1/\mu = -\mathbf{p}_2/\mu$).
It reads
\be
\label{eq5.9}
H_{\rm 4PN}[\mathbf{r},\mathbf{p}] = H_{\rm 4PN}^{\rm loc}(\mathbf{r},\mathbf{p}) + H_{\rm 4PN}^{\rm nonloc} \,,
\ee
where the nonlocal piece can be written as
\begin{align}
\label{eq5.10}
H_{{\rm 4PN}}^{\rm nonloc}(t) &= -\frac15 \, \frac{G^2 M}{c^8} \, I_{ij}^{(3)}(t)
\nonumber\\[1ex]
&\qquad\times\,\Pf_{2r_{12}/c} \int_{-\infty}^{+\infty} \frac{\md v}{\vert v \vert} \, I_{ij}^{(3)} (t+v) \, ,
\end{align}
and where the final local piece, $H_{\rm 4PN}^{\rm loc} = H_{\rm 4PN}^{\rm loc \, 0} + C F [\mathbf{r},\mathbf{p}]$,
incorporating the value (\ref{eq5.8}) of $C$, is explicitly given by
\begin{widetext}
\begin{align}
\label{eq5.11}
&c^8\,\widehat{H}_{\rm 4PN}^{\rm loc}(\mathbf{r},\mathbf{p}) := c^8\,\frac{H_{\rm 4PN}^{\rm loc}(\mathbf{r},\mathbf{p})}\mu = 
\bigg( \frac{7}{256} - \frac{63}{256}\nu +\frac{189}{256}\nu^2 - \frac{105}{128}\nu^3 + \frac{63}{256}\nu^4 \bigg) \ppn^5
\nonumber\\[1ex]&\quad
+ \Bigg\{
\frac{45}{128}\ppn^4 - \frac{45}{16}\ppn^4\,\nu
+\left( \frac{423}{64}\ppn^4 -\frac{3}{32}\np^2\ppn^3 - \frac{9}{64}\np^4\ppn^2 \right)\,\nu^2
\nonumber\\[1ex]&\quad
+ \left( -\frac{1013}{256}\ppn^4 + \frac{23}{64}\np^2\ppn^3 + \frac{69}{128}\np^4\ppn^2
- \frac{5}{64}\np^6\pp + \frac{35}{256}\np^8 \right)\,\nu^3
\nonumber\\[1ex]&\quad
+ \left( -\frac{35}{128}\ppn^4 - \frac{5}{32}\np^2\ppn^3 - \frac{9}{64}\np^4\ppn^2 -\frac{5}{32}\np^6\pp
- \frac{35}{128}\np^8 \right)\,\nu^4
\Bigg\}\frac{1}{r}
\nonumber\\[1ex]&\quad
+ \Bigg\{ \frac{13}{8}\ppn^3
+ \left( -\frac{791}{64}\ppn^3 + \frac{49}{16}\np^2\ppn^2 -\frac{889}{192}\np^4\pp + \frac{369}{160}\np^6 \right)\,\nu
\nonumber\\[1ex]&\quad
+\left( \frac{4857}{256}\ppn^3 -\frac{545}{64}\np^2\ppn^2 +\frac{9475}{768}\np^4\pp - \frac{1151}{128}\np^6 \right)\,\nu^2
\nonumber\\[1ex]&\quad
+ \left( \frac{2335}{256}\ppn^3 + \frac{1135}{256}\np^2\ppn^2 - \frac{1649}{768}\np^4\pp + \frac{10353}{1280}\np^6 \right)\,\nu^3
\Bigg\}\frac{1}{r^2}
\nonumber\\[1ex]&\quad
+ \Bigg\{ \frac{105}{32}\ppn^2
+ \bigglb( \left(\frac{2749\pi^2}{8192}-\frac{589189}{19200}\right)\ppn^2
+ \left(\frac{63347}{1600}-\frac{1059\pi^2}{1024}\right)\np^2\pp
+ \left(\frac{375\pi^2}{8192}-\frac{23533}{1280}\right)\np^4 \biggrb)\,\nu
\nonumber\\[1ex]&\quad
+ \bigglb( \left(\frac{18491\pi^2}{16384}-\frac{1189789}{28800}\right)\ppn^2
+ \left(-\frac{127}{3}-\frac{4035\pi^2}{2048}\right)\np^2\pp
+ \left(\frac{57563}{1920}-\frac{38655\pi^2}{16384}\right)\np^4 \biggrb)\,\nu^2
\nonumber\\[1ex]&\quad
+\left( -\frac{553}{128}\ppn^2 -\frac{225}{64}\np^2\pp -\frac{381}{128}\np^4 \right)\,\nu^3
\Bigg\}\frac{1}{r^3}
\nonumber\\[1ex]&\quad
+ \Bigg\{
\frac{105}{32}\pp
+ \bigglb( \left(\frac{185761}{19200}-\frac{21837\pi^2}{8192}\right) \pp
+ \left(\frac{3401779}{57600}-\frac{28691\pi^2}{24576}\right) \np^2 \biggrb)\,\nu
\nonumber\\[1ex]&\quad
+ \bigglb( \left(\frac{672811}{19200}-\frac{158177\pi^2}{49152}\right) \pp
+ \left(\frac{110099\pi^2}{49152}-\frac{21827}{3840}\right) \np^2 \biggrb)\,\nu^2 \Bigg\}\frac{1}{r^4}
\nonumber\\[1ex]&\quad
+ \Bigg\{ -\frac{1}{16} + \left(\frac{6237\pi^2}{1024}-\frac{169199}{2400}\right)\,\nu 
+ \left(\frac{7403\pi^2}{3072}-\frac{1256}{45} \right)\,\nu^2 \Bigg\}\frac{1}{r^5}.
\end{align}
\end{widetext}
Here, we have reduced the order of time derivatives,
which is allowed modulo suitable variable redefinitions \cite{Schafer84,DS1985,DS1991}.
Concerning the time derivatives entering the nonlocal piece (\ref{eq5.10}) via $I_{ij}^{(3)}(t)$,
we can either consider that $I_{ij}^{(3)}$ is defined as the third time derivative of the center-of-mass quadrupole moment
\begin{align}
I_{ij} &= \mu \left( x_{12}^i \, x_{12}^j - \frac13 \, \mathbf{x}_{12}^2 \, \delta^{ij} \right)
\nonumber\\[1ex]
&= (GM)^2 \mu \left( r^i \, r^j - \frac13 \,  \mathbf{r}^2 \, \delta^{ij} \right)
\end{align}
and does depend on $\mathbf{r}$, $\dot{\mathbf{r}}$, $\ddot{\mathbf{r}}$, and $\dddot{\mathbf{r}}$
[which might be convenient for integrating (\ref{eq5.10}) by parts], or define (\ref{eq5.10})
by inserting in it the order-reduced value of $I_{ij}^{(3)}$,
namely (at the Newtonian accuracy, which is sufficient)
\begin{align}
\label{eq5.12}
\left(I_{ij}^{(3)}\right)_\textrm{red} &= -2\,\frac{G\mu\,M}{r_{12}^3}
\nonumber\\[1ex]&\quad\times\,
\left(4 \, x_{12}^{\langle i} v_{12}^{j\rangle}
- \frac3{r_{12}} \, (\mathbf{n}_{12} \cdot \mathbf{v}_{12}) \, x_{12}^{\langle i} \, x_{12}^{j\rangle} \right).
\end{align}
Here $\langle{ij}\rangle$ denotes a symmetric tracefree projection.
This Galileo-invariant result (with $\mathbf{v}_{12}:=\mathbf{p}_1/m_1-\mathbf{p}_2/m_2$) is valid in an arbitrary frame.
In the center-of-mass frame one only needs to interpret $\mathbf{x}_{12}$ as $GM\,\mathbf{r}$ and $\mathbf{v}_{12}$ as $\mathbf{p}$.

A partially explicit, partially parametrized formula for the 4PN local Hamiltonian in the center-of-mass frame
was given in Ref.\ \cite{Jaranowski:2013lca} [see Eqs.\ (4.2)--(4.4) there].
The comparison of this formula with our Eq.\ \eqref{eq5.11} shows
that all the numerical coefficients explicitly displayed in \cite{Jaranowski:2013lca}
coincide with the corresponding coefficients in Eq.\ \eqref{eq5.11}.
Moreover, the values of the six coefficients that were not determined in Ref.\ \cite{Jaranowski:2013lca}
(namely $c_{411}$, $c_{412}$, $c_{413}$, $c_{211}$, $c_{212}$, $c_{01}$), can now be read off from Eq.\ \eqref{eq5.11}.
The logarithmic terms $\propto \ln r/\hat s$ displayed in Eq.\ (4.2) of Ref.\ \cite{Jaranowski:2013lca} correspond
(modulo a total time derivative) to the term $F \ln r_{12} / s$ in Eq.\ \eqref{eq3.6} with $F$
defined by Eq.\ \eqref{eq3.7}, with
$$
G^2 \big(I_{ij}^{(3)}\big)^2(\mathbf{r},\mathbf{p}) = \frac83 \, \nu^2 \, \frac1{r^4} \,
(12 \mathbf{p}^2 - 11 (\mathbf{n} \cdot \mathbf{p})^2) \,.
$$

For the convenience of the reader, let us complete the 4PN-level Hamiltonian obtained above
by the explicit center-of-mass expressions of the lower-PN-levels Hamiltonians.
The 4PN-accurate reduced center-of-mass Hamiltonian
$\widehat{H}_\textrm{$\le$4PN}:=({H}_\textrm{$\le$4PN}-M c^2)/\mu$ reads
\begin{align}
\widehat{H}_\textrm{$\le$4PN}[\mathbf{r},\mathbf{p}]
&= \widehat{H}_{\rm N}(\mathbf{r},\mathbf{p})+ \widehat{H}_{\rm 1PN}(\mathbf{r},\mathbf{p})
+ \widehat{H}_{\rm 2PN}(\mathbf{r},\mathbf{p})
\nonumber\\[1ex]&\quad
+ \widehat{H}_{\rm 3PN}(\mathbf{r},\mathbf{p}) + \widehat{H}_{\rm 4PN}[\mathbf{r},\mathbf{p}],
\end{align}
where the Hamiltonian $\widehat{H}_{\rm 4PN}$ is determined by Eqs.\ \eqref{eq5.9}--\eqref{eq5.11}
and the (purely local) Hamiltonians $\widehat{H}_{\rm N}$ to $\widehat{H}_{\rm 3PN}$ are equal to
\begin{widetext}
\begin{subequations}
\begin{align}
\widehat{H}_{\rm N}\left({\bf r},{\bf p}\right) &= \frac{\pp}{2} - \frac{1}{r},
\\[2ex]
c^2\,\widehat{H}_{\rm 1PN}\left({\bf r},{\bf p}\right) &=
\frac{1}{8}(3\nu-1) \ppn^2 - \frac{1}{2}\Big\{(3+\nu)\pp + \nu \np^2
\Big\}\frac{1}{r} + \frac{1}{2r^2},
\\[2ex]
c^4\,\widehat{H}_{\rm 2PN}\left({\bf r},{\bf p}\right) &= 
\frac{1}{16}\left(1-5\nu+5\nu^2\right) \ppn^3
+ \frac{1}{8}\Big\{ \left(5-20\nu-3\nu^2\right)\ppn^2 - 2\nu^2 \np^2\pp - 3\nu^2\np^4 \Big\}\frac{1}{r}
\nonumber\\[2ex]&\quad
+ \frac{1}{2} \Big\{(5+8\nu)\pp+3\nu\np^2\Big\}\frac{1}{r^2}
- \frac{1}{4}(1+3\nu)\frac{1}{r^3},
\\[2ex]
c^6\,\widehat{H}_{\rm 3PN}\left({\bf r},{\bf p}\right)
&= \frac{1}{128}\left(-5+35\nu-70\nu^2+35\nu^3\right)\ppn^4
+ \frac{1}{16}\bigg\{
\left(-7+42\nu-53\nu^2-5\nu^3\right)\ppn^3
\nonumber\\[2ex]&\quad
+ (2-3\nu)\nu^2\np^2\ppn^2
+ 3(1-\nu)\nu^2\np^4\pp - 5\nu^3\np^6
\bigg\}\frac{1}{r}
\nonumber\\[2ex]&\quad
+\bigg\{ \frac{1}{16}\left(-27+136\nu+109\nu^2\right)\ppn^2
+ \frac{1}{16}(17+30\nu)\nu\np^2\pp + \frac{1}{12}(5+43\nu)\nu\np^4
\bigg\}\frac{1}{r^2}
\nonumber\\[2ex]&\quad
+\Bigg\{ \bigglb(
-\frac{25}{8}+\left(\frac{\pi^2}{64}-\frac{335}{48}\right)\nu-\frac{23\nu^2}{8} \biggrb)\pp
+ \left(-\frac{85}{16}-\frac{3\pi^2}{64}-\frac{7\nu}{4}\right)\nu\np^2 
\Bigg\}\frac{1}{r^3}
\nonumber\\[2ex]&\quad
+ \Bigg\{\frac{1}{8}+\Big(\frac{109}{12}-\frac{21}{32}\pi^2\Big)\nu\Bigg\}\frac{1}{r^4}.
\end{align}
\end{subequations}
\end{widetext}

\section{Discussion}

The results presented here complete a line of work which has been started years ago.
The most striking new feature of our result (\ref{eq5.9}) for the 4PN-accurate action is its explicit time-symmetric nonlocality in time.
This nonlocality was to be a priori expected in view of the structure of the higher-order contributions to the action, such as Eq.\ (\ref{eq2.2}).
Note, however, that the explicit calculation of the (gradient of) the $O(G^2)$ action
[Eq.\ (\ref{eq2.2}), together with other terms] in Ref.\ \cite{Bel1981},
using straight (nonparallel) world lines, did not give rise to any IR problem,
and led to a final result expressed in terms of (quasilocal) lightlike related quantities on the two world lines.
In the ADM approach, the nonlocality of the action is more concentrated than in the harmonic-gauge action
because the PN most prominent interactions mediated by the scalar ($\phi$) and vector ($V^i$) degrees of freedom are both instantaneous.
The time-nonlocality is only due to the interaction mediated by $h^{\rm TT}_{ij}$, which starts contributing at the 2PN level.
At the 4PN level one can no longer approximate the propagator of $h^{\rm TT}_{ij}$ by a PN expansion of the type (\ref{eq1.1})
because one must take into account the backscatter of the $h^{\rm TT}_{ij}$ propagator
due to its coupling to the space curvature linked with $\phi_{(2)}$.

Note, however, that, when completing the equations of motion deriving from Eq.\ (\ref{eq5.9}) [or Eq.\ (\ref{eq5.1})]
by adding the corresponding 4PN-level radiation-reaction force,
acting on each body $(\dot p_a^i = {\mathcal F}_a^{i \, {\rm sym}} + {\mathcal F}_a^{i \, {\rm rad \, reac}})$,
namely, from Eq.\ (\ref{eq4.3})
\begin{multline*}
{\mathcal F}_{a \, {\rm 4PN}}^{i \,{\rm rad \, reac}}
= + \frac{1}{2} m_a c^2 \left(\partial_i \, h_{\rm 00 \, 4PN}^{\rm rad \, reac} \right)_{x^i = x_a^i}\quad
\\[1ex]
= -\frac45 \, \frac{G^2 \, M}{c^8} \, m_a x_a^j \int_0^{+\infty} \frac{\md v}v \left( I_{ij}^{(6)} (t-v) -I_{ij}^{(6)} (t+v) \right),
\end{multline*}
the ``advanced'' piece (with time argument $t+v$) in this time-antisymmetric radiation reaction
precisely cancels a corresponding advanced contribution in the time-symmetric conservative force
deriving from the nonlocal piece in Eq.~(\ref{eq5.1}).
[This cancellation is also clear in Eqs.\ (\ref{eqs4.1-4.3}).]

Our results open new avenues for further investigations.
Here, we have only considered the consequences of our action for the dynamics of circular orbits.
For these orbits, our nonlocal action induces an ordinary, local radial potential $A(r)$ when cast within the effective one-body framework.
We leave to future work the full recasting of our action within the effective one-body formalism. 
Our work also indicates that the further logarithmic contributions entering the radial potential $A(r)$ beyond the
4PN level \cite{Blanchet:2010,Damourlogs,LeTiec:2012,Barausse:2011dq,Bini:2013rfa} will all be associated with 
nonlocal actions of the type  of Eq.~(\ref{eq4.5}) (involving higher multipoles, or PN corrections to the quadrupole).
(This is clear from the derivation in \cite{Bini:2013rfa} where one sees that these logarithms 
are associated with hereditary contributions in the inner metric linked with tail effects in higher multipoles.)

\begin{acknowledgments}

P.J.\ gratefully acknowledges support of the Deutsche Forschungsgemeinschaft (DFG)
through the Transregional Collaborative Research Center SFB/TR7
``Gravitational Wave Astronomy: Methods--Sources--Observation.''
P.J.\ also thanks the Institut des Hautes \'Etudes Scientifiques
for hospitality during the crucial stage of the collaboration.
The work of P.J.\ was supported in part by the Polish MNiSW Grant No.\ 623/N-VIRGO/09/2010/0.

\end{acknowledgments}

\appendix

\section{Noncenter-of-mass 4PN-accurate Hamiltonian and the boost vector}

In this appendix we show, for completeness and convenience of the reader,
the generic (i.e., noncenter-of-mass) form of the 4PN-accurate \emph{local} Hamiltonian
$H_\textrm{$\le$4PN}^\textrm{local}$. It reads
\begin{align}
\label{ah4pn}
H_\textrm{$\le$4PN}^\textrm{local}(\mathbf{x}_a,\mathbf{p}_a)
&= Mc^2 + H_\textrm{N}(\mathbf{x}_a,\mathbf{p}_a) + H_\textrm{1PN}(\mathbf{x}_a,\mathbf{p}_a)
\nonumber\\[1.5ex]&\quad
+ H_\textrm{2PN}(\mathbf{x}_a,\mathbf{p}_a) + H_\textrm{3PN}(\mathbf{x}_a,\mathbf{p}_a)
\nonumber\\[1.5ex]&\quad
+ H_\textrm{4PN}^\textrm{local}(\mathbf{x}_a,\mathbf{p}_a),
\end{align}
where the 4PN local piece,
$H_\textrm{4PN}^\textrm{local}(\mathbf{x}_a,\mathbf{p}_a)
=H_\textrm{4PN}^\textrm{local\,0}(\mathbf{x}_a,\mathbf{p}_a)+C F(\mathbf{x}_a,\mathbf{p}_a)$,
incorporates the value \eqref{eq5.8} of the constant $C$.
[On the other hand, it does not contain any logarithmic contribution $\propto F(\mathbf{x}_a,\mathbf{p}_a)\ln(r_{12}/s)$;
indeed, we have incorporated these logarithmic contributions in the definition \eqref{eq5.10}
of the complementary nonlocal 4PN Hamiltonian, as per Eq.\ \eqref{eq5.9}].

The Hamiltonians $H_\textrm{N}$ to $H_\textrm{3PN}$ are equal to
[the operation ``$+\big(1\leftrightarrow 2\big)$'' used below denotes the addition for each term,
including the ones which are symmetric under the exchange of body labels,
of another term obtained by the label permutation $1\leftrightarrow 2$]
\begin{widetext}
\begin{subequations}
\begin{align}
H_{\text{N}}({\bf x}_a,{\bf p}_a) &= \frac{{\bf p}_1^2}{2\,m_1}
- \frac{1}{2}\frac{G \, m_1 \, m_2}{r_{12}}
+ \big(1\leftrightarrow 2\big),
\\[2ex]
c^2\,H_{\text{1PN}}({\bf x}_a,{\bf p}_a) &=
- \frac{1}{8}\frac{\pipip^2}{m_1^3}
+ \frac{1}{8}\frac{Gm_1m_2}{r_{12}} \left(
- 12\,\frac{\pipi}{m_1^2}
+ 14\,\frac{\pipj}{m_1m_2}
+ 2\,\frac{\npi\npj}{m_1m_2} \right)
\nonumber\\[2ex]&\quad
+ \frac{1}{4}\frac{Gm_1m_2}{r_{12}}\frac{G(m_1+m_2)}{r_{12}}
+ \big(1\leftrightarrow 2\big),
\\[2ex]
c^4\,H_{\text{2PN}}({\bf x}_a,{\bf p}_a) &=
\frac{1}{16}\frac{\pipip^3}{m_1^5}
+ \frac{1}{8} \frac{Gm_1m_2}{r_{12}} \Bigg(
5\,\frac{\pipip^2}{m_1^4}
- \frac{11}{2}\frac{\pipi\,\pjpj}{m_1^2m_2^2}
- \frac{\pipj^2}{m_1^2m_2^2}
+ 5\,\frac{\pipi\,\npj^2}{m_1^2m_2^2}
\nonumber\\[2ex]&\quad
- 6\,\frac{\pipj\,\npi\npj}{m_1^2m_2^2}
- \frac{3}{2}\frac{\npi^2\npj^2}{m_1^2m_2^2} 
\Bigg)
\nonumber\\[2ex]&\quad
+ \frac{1}{4}\frac{G^2m_1m_2}{r_{12}^2} \Bigglb(
m_2\left(10\frac{\pipi}{m_1^2}+19\frac{\pjpj}{m_2^2}\right)
- \frac{1}{2}(m_1+m_2)\frac{27\,\pipj+6\,\npi\npj}{m_1m_2} \Biggrb)
\nonumber\\[2ex]&\quad
- \frac{1}{8} \frac{Gm_1m_2}{r_{12}}\frac{G^2 (m_1^2+5m_1m_2+m_2^2)}{r_{12}^2}
+ \big(1\leftrightarrow 2\big),
\\[2ex]
c^6\,H_{\text{3PN}}({\bf x}_a,{\bf p}_a) &=
-\frac{5}{128}\frac{\pipip^4}{m_1^7}
+ \frac{1}{32} \frac{Gm_1m_2}{r_{12}} \Bigg(
- 14\,\frac{\pipip^3}{m_1^6}
+ 4\,\frac{\biglb(\pipj^2+4\,\pipi\,\pjpj\bigrb)\pipi}{m_1^4m_2^2}
+ 6\,\frac{\pipi\,\npi^2\npj^2}{m_1^4m_2^2}
\nonumber\\[2ex]&\quad
- 10\,\frac{\biglb(\pipi\,\npj^2+\pjpj\,\npi^2\bigrb)\pipi}{m_1^4m_2^2}
+ 24\,\frac{\pipi\,\pipj\npi\npj}{m_1^4m_2^2}
\nonumber\\[2ex]&\quad
+ 2\,\frac{\pipi\,\pipj\npj^2}{m_1^3m_2^3}
+ \frac{\biglb(7\,\pipi\,\pjpj-10\,\pipj^2\bigrb)\npi\npj}{m_1^3m_2^3}
\nonumber\\[2ex]&\quad
+ \frac{\biglb(\pipi\,\pjpj-2\,\pipj^2\bigrb)\pipj}{m_1^3m_2^3}
+ 15\,\frac{\pipj\npi^2\npj^2}{m_1^3m_2^3}
\nonumber\\[2ex]&\quad
- 18\,\frac{\pipi\,\npi\npj^3}{m_1^3m_2^3}
+ 5\,\frac{\npi^3\npj^3}{m_1^3m_2^3} \Bigg)
+ \frac{G^2m_1m_2}{r_{12}^2} \Bigg(
\frac{1}{16}(m_1-27m_2)\frac{\pipip^2}{m_1^4}
\nonumber\\[2ex]&\quad
- \frac{115}{16}m_1\frac{\pipi\,\pipj}{m_1^3m_2}
+ \frac{1}{48}m_2\frac{25\,\pipj^2+371\,\pipi\,\pjpj}{m_1^2 m_2^2}
+ \frac{17}{16}\frac{\pipi\npi^2}{m_1^3}
+ \frac{5}{12}\frac{\npi^4}{m_1^3}
\nonumber\\[2ex]&\quad
- \frac{1}{8}m_1 \frac{\biglb(15\,\pipi\,\npj+11\,\pipj\,\npi\bigrb)\npi}{m_1^3 m_2}
- \frac{3}{2}m_1\frac{\npi^3\npj}{m_1^3m_2}
\nonumber\\[2ex]&\quad
+ \frac{125}{12}m_2\frac{\pipj\,\npi\npj}{m_1^2m_2^2}
+ \frac{10}{3}m_2\frac{\npi^2\npj^2}{m_1^2m_2^2}
\nonumber\\[2ex]&\quad
- \frac{1}{48} (220 m_1 + 193 m_2) \frac{\pipi \npj^2}{m_1^2 m_2^2}
\Bigg)
\nonumber\\[2ex]&\quad
+ \frac{G^3m_1m_2}{r_{12}^3} \Bigg(
-\frac{1}{48}
\bigglb(425\,m_1^2+\Big(473-\frac{3}{4}\pi^2\Big)m_1m_2+150\,m_2^2\biggrb)
\frac{\pipi}{m_1^2}
\nonumber\\[2ex]&\quad
+ \frac{1}{16}
\bigglb(77(m_1^2+m_2^2)+\Big(143-\frac{1}{4}\pi^2\Big)m_1m_2\biggrb)
\frac{\pipj}{m_1m_2}
\nonumber\\[2ex]&\quad
+ \frac{1}{16}
\bigglb(20\,m_1^2-\Big(43+\frac{3}{4}\pi^2\Big)m_1m_2\biggrb)
\frac{\npi^2}{m_1^2}
\nonumber\\[2ex]&\quad
+ \frac{1}{16}
\bigglb(21(m_1^2+m_2^2)+\Big(119+\frac{3}{4}\pi^2\Big)m_1m_2\biggrb)
\frac{\npi\npj}{m_1m_2} \Bigg)
\nonumber\\[2ex]&\quad
+ \frac{1}{8} \frac{G^4m_1m_2^3}{r_{12}^4}
\Bigg( \bigg(\frac{227}{3}-\frac{21}{4}\pi^2\bigg)m_1+m_2 \Bigg)
+ \big(1\leftrightarrow 2\big).
\end{align}
\end{subequations}

The formula for the Hamiltonian $H_\textrm{4PN}^\textrm{local}$ is very large,
therefore we display it in smaller pieces. This Hamiltonian has the following structure
\begin{align}
c^8\,H_\textrm{4PN}^\textrm{local}(\mathbf{x}_a,\mathbf{p}_a) &=
\frac{7 \pipip^5}{256 m_1^9}
+ \frac{G m_1 m_2}{r_{12}} H_{48}(\mathbf{x}_a,\mathbf{p}_a)
+ \frac{G^2 m_1 m_2}{r_{12}^2} m_1\,H_{46}(\mathbf{x}_a,\mathbf{p}_a)
\nonumber\\[1ex]&\quad
+ \frac{G^3 m_1 m_2}{r_{12}^3} \Big(m_1^2\,H_{441}(\mathbf{x}_a,\mathbf{p}_a) + m_1 m_2\,H_{442}(\mathbf{x}_a,\mathbf{p}_a) \Big)
\nonumber\\[1ex]&\quad
+ \frac{G^4 m_1 m_2}{r_{12}^4} \Big(m_1^3\,H_{421}(\mathbf{x}_a,\mathbf{p}_a) + m_1^2 m_2\,H_{422}(\mathbf{x}_a,\mathbf{p}_a)\Big)
+ \frac{G^5 m_1 m_2}{r_{12}^5} H_{40}(\mathbf{x}_a,\mathbf{p}_a)
+ \big(1\leftrightarrow 2\big),
\end{align}
where
\begin{subequations}
\begin{align}
H_{48}(\mathbf{x}_a,\mathbf{p}_a) &=\frac{45 \pipip^4}{128 m_1^8}
-\frac{9 \npi^2 \npii^2 \pipip^2}{64 m_1^6 m_2^2}
+\frac{15 \npii^2 \pipip^3}{64m_1^6 m_2^2}
-\frac{9 \npi \npii \pipip^2 \pipii}{16 m_1^6 m_2^2}
\nonumber\\[1ex]&\quad
-\frac{3 \pipip^2 \pipii^2}{32m_1^6 m_2^2}
+\frac{15 \npi^2 \pipip^2 \piipii}{64 m_1^6 m_2^2}
-\frac{21 \pipip^3 \piipii}{64 m_1^6m_2^2}
-\frac{35 \npi^5 \npii^3}{256 m_1^5 m_2^3}
\nonumber\\[1ex]&\quad
+\frac{25 \npi^3 \npii^3 \pipi}{128 m_1^5m_2^3}
+\frac{33 \npi \npii^3 \pipip^2}{256 m_1^5 m_2^3}
-\frac{85 \npi^4 \npii^2 \pipii}{256 m_1^5m_2^3}
\nonumber\\[1ex]&\quad
-\frac{45 \npi^2 \npii^2 \pipi \pipii}{128 m_1^5 m_2^3}
-\frac{\npii^2 \pipip^2 \pipii}{256m_1^5 m_2^3}
+\frac{25 \npi^3 \npii \pipii^2}{64 m_1^5 m_2^3}
\nonumber\\[1ex]&\quad
+\frac{7 \npi \npii \pipi\pipii^2}{64 m_1^5 m_2^3}
-\frac{3 \npi^2 \pipii^3}{64 m_1^5 m_2^3}
+\frac{3 \pipi \pipii^3}{64 m_1^5m_2^3}
\nonumber\\[1ex]&\quad
+\frac{55 \npi^5 \npii \piipii}{256 m_1^5 m_2^3}
-\frac{7 \npi^3 \npii \pipi \piipii}{128m_1^5 m_2^3}
-\frac{25 \npi \npii \pipip^2 \piipii}{256 m_1^5 m_2^3}
\nonumber\\[1ex]&\quad
-\frac{23 \npi^4 \pipii\piipii}{256 m_1^5 m_2^3}
+\frac{7 \npi^2 \pipi \pipii \piipii}{128 m_1^5 m_2^3}
-\frac{7 \pipip^2\pipii \piipii}{256 m_1^5 m_2^3}
\nonumber\\[1ex]&\quad
-\frac{5 \npi^2 \npii^4 \pipi}{64 m_1^4 m_2^4}
+\frac{7 \npii^4\pipip^2}{64 m_1^4 m_2^4}
-\frac{\npi \npii^3 \pipi \pipii}{4 m_1^4 m_2^4}
\nonumber\\[1ex]&\quad
+\frac{\npii^2 \pipi\pipii^2}{16 m_1^4 m_2^4}
-\frac{5 \npi^4 \npii^2 \piipii}{64 m_1^4 m_2^4}
+\frac{21 \npi^2 \npii^2\pipi \piipii}{64 m_1^4 m_2^4}
\nonumber\\[1ex]&\quad
-\frac{3 \npii^2 \pipip^2 \piipii}{32 m_1^4 m_2^4}
-\frac{\npi^3\npii \pipii \piipii}{4 m_1^4 m_2^4}
+\frac{\npi \npii \pipi \pipii \piipii}{16 m_1^4m_2^4}
\nonumber\\[1ex]&\quad
+\frac{\npi^2 \pipii^2 \piipii}{16 m_1^4 m_2^4}
-\frac{\pipi \pipii^2 \piipii}{32 m_1^4m_2^4}
+\frac{7 \npi^4 \piipiip^2}{64 m_1^4 m_2^4}
\nonumber\\[1ex]&\quad
-\frac{3 \npi^2 \pipi \piipiip^2}{32 m_1^4m_2^4}
-\frac{7 \pipip^2 \piipiip^2}{128 m_1^4 m_2^4},
\\[3ex]
H_{46}(\mathbf{x}_a,\mathbf{p}_a) &=
\frac{369 \npi^6}{160 m_1^6}
-\frac{889 \npi^4 \pipi}{192 m_1^6}
+\frac{49 \npi^2 \pipip^2}{16 m_1^6}
-\frac{63\pipip^3}{64 m_1^6}
\nonumber\\[1ex]&\quad
-\frac{549 \npi^5 \npii}{128 m_1^5 m_2}
+\frac{67 \npi^3 \npii \pipi}{16 m_1^5m_2}
-\frac{167 \npi \npii \pipip^2}{128 m_1^5 m_2}
\nonumber\\[1ex]&\quad
+\frac{1547 \npi^4 \pipii}{256 m_1^5m_2}
-\frac{851 \npi^2 \pipi \pipii}{128 m_1^5 m_2}
+\frac{1099 \pipip^2 \pipii}{256 m_1^5 m_2}
\nonumber\\[1ex]&\quad
+\frac{3263\npi^4 \npii^2}{1280 m_1^4 m_2^2}
+\frac{1067 \npi^2 \npii^2 \pipi}{480 m_1^4 m_2^2}
-\frac{4567\npii^2 \pipip^2}{3840 m_1^4 m_2^2}
\nonumber\\[1ex]&\quad
-\frac{3571 \npi^3 \npii \pipii}{320 m_1^4 m_2^2}
+\frac{3073\npi \npii \pipi \pipii}{480 m_1^4 m_2^2}
+\frac{4349 \npi^2 \pipii^2}{1280 m_1^4 m_2^2}
\nonumber\\[1ex]&\quad
-\frac{3461\pipi \pipii^2}{3840 m_1^4 m_2^2}
+\frac{1673 \npi^4 \piipii}{1920 m_1^4 m_2^2}
-\frac{1999 \npi^2 \pipi\piipii}{3840 m_1^4 m_2^2}
\nonumber\\[1ex]&\quad
+\frac{2081 \pipip^2 \piipii}{3840 m_1^4 m_2^2}
-\frac{13 \npi^3 \npii^3}{8m_1^3 m_2^3}
+\frac{191 \npi \npii^3 \pipi}{192 m_1^3 m_2^3}
\nonumber\\[1ex]&\quad
-\frac{19 \npi^2 \npii^2 \pipii}{384m_1^3 m_2^3}
-\frac{5 \npii^2 \pipi \pipii}{384 m_1^3 m_2^3}
+\frac{11 \npi \npii \pipii^2}{192m_1^3 m_2^3}
\nonumber\\[1ex]&\quad
+\frac{77 \pipii^3}{96 m_1^3 m_2^3}
+\frac{233 \npi^3 \npii \piipii}{96 m_1^3m_2^3}
-\frac{47 \npi \npii \pipi \piipii}{32 m_1^3 m_2^3}
\nonumber\\[1ex]&\quad
+\frac{\npi^2 \pipii \piipii}{384 m_1^3m_2^3}
-\frac{185 \pipi \pipii \piipii}{384 m_1^3 m_2^3}
-\frac{7 \npi^2 \npii^4}{4 m_1^2 m_2^4}
\nonumber\\[1ex]&\quad
+\frac{7\npii^4 \pipi}{4 m_1^2 m_2^4}
-\frac{7 \npi \npii^3 \pipii}{2 m_1^2 m_2^4}
+\frac{21 \npii^2\pipii^2}{16 m_1^2 m_2^4}
\nonumber\\[1ex]&\quad
+\frac{7 \npi^2 \npii^2 \piipii}{6 m_1^2 m_2^4}
+\frac{49 \npii^2 \pipi\piipii}{48 m_1^2 m_2^4}
-\frac{133 \npi \npii \pipii \piipii}{24 m_1^2 m_2^4}
\nonumber\\[1ex]&\quad
-\frac{77 \pipii^2\piipii}{96 m_1^2 m_2^4}
+\frac{197 \npi^2 \piipiip^2}{96 m_1^2 m_2^4}
-\frac{173 \pipi \piipiip^2}{48 m_1^2m_2^4}
+\frac{13 \piipiip^3}{8 m_2^6},
\\[3ex]
H_{441}(\mathbf{x}_a,\mathbf{p}_a) &=
\frac{5027 \npi^4}{384 m_1^4}
-\frac{22993 \npi^2 \pipi}{960 m_1^4}
-\frac{6695 \pipip^2}{1152 m_1^4}
-\frac{3191\npi^3 \npii}{640 m_1^3 m_2}
\nonumber\\[1ex]&\quad
+\frac{28561 \npi \npii \pipi}{1920 m_1^3 m_2}
+\frac{8777 \npi^2\pipii}{384 m_1^3 m_2}
+\frac{752969 \pipi \pipii}{28800 m_1^3 m_2}
\nonumber\\[1ex]&\quad
-\frac{16481 \npi^2 \npii^2}{960m_1^2 m_2^2}
+\frac{94433 \npii^2 \pipi}{4800 m_1^2 m_2^2}
-\frac{103957 \npi \npii \pipii}{2400 m_1^2m_2^2}
\nonumber\\[1ex]&\quad
+\frac{791 \pipii^2}{400 m_1^2 m_2^2}
+\frac{26627 \npi^2 \piipii}{1600 m_1^2 m_2^2}
-\frac{118261 \pipi\piipii}{4800 m_1^2 m_2^2}
+\frac{105 \piipiip^2}{32 m_2^4},
\\[3ex]
H_{442}(\mathbf{x}_a,\mathbf{p}_a) &=
\left(\frac{2749 \pi ^2}{8192}-\frac{211189}{19200}\right) \frac{\pipip^2}{m_1^4}
+\left(\frac{63347}{1600}-\frac{1059 \pi ^2}{1024}\right)\frac{\npi^2 \pipi}{m_1^4}
+\left(\frac{375\pi^2}{8192}-\frac{23533}{1280}\right)\frac{\npi^4}{m_1^4}
\nonumber\\[1ex]&\quad
+\left(\frac{10631 \pi ^2}{8192}-\frac{1918349}{57600}\right) \frac{\pipii^2}{m_1^2 m_2^2}
+\left(\frac{13723 \pi^2}{16384}-\frac{2492417}{57600}\right) \frac{\pipi\piipii}{m_1^2 m_2^2}
\nonumber\\[1ex]&\quad
+\left(\frac{1411429}{19200}-\frac{1059 \pi ^2}{512}\right)\frac{\npii^2\pipi}{m_1^2 m_2^2}
+\left(\frac{248991}{6400}-\frac{6153 \pi ^2}{2048}\right) \frac{\npi\npii\pipii}{m_1^2 m_2^2}
\nonumber\\[1ex]&\quad
-\left(\frac{30383}{960}+\frac{36405\pi^2}{16384}\right) \frac{\npi^2 \npii^2}{m_1^2m_2^2}
+\left(\frac{1243717}{14400}-\frac{40483 \pi ^2}{16384}\right) \frac{\pipi\pipii}{m_1^3m_2}
\nonumber\\[1ex]&\quad
+\left(\frac{2369}{60}+\frac{35655 \pi ^2}{16384}\right) \frac{\npi^3 \npii}{m_1^3 m_2}
+\left(\frac{43101\pi ^2}{16384}-\frac{391711}{6400}\right) \frac{\npi \npii \pipi}{m_1^3 m_2}
\nonumber\\[1ex]&\quad
+\left(\frac{56955 \pi^2}{16384}-\frac{1646983}{19200}\right) \frac{\npi^2 \pipii}{m_1^3 m_2},
\\[3ex]
H_{421}(\mathbf{x}_a,\mathbf{p}_a) &=
\frac{64861 \pipi}{4800 m_1^2}-\frac{91 \pipii}{8 m_1 m_2}
+\frac{105 \piipii}{32 m_2^2}
-\frac{9841 \npi^2}{1600m_1^2}-\frac{7 \npi \npii}{2 m_1 m_2},
\\[3ex]
H_{422}(\mathbf{x}_a,\mathbf{p}_a) &=
\left(\frac{1937033}{57600}-\frac{199177 \pi ^2}{49152}\right) \frac{\pipi}{m_1^2}
+\left(\frac{176033 \pi ^2}{24576}-\frac{2864917}{57600}\right)\frac{\pipii}{m_1 m_2}
+\left(\frac{282361}{19200}-\frac{21837 \pi ^2}{8192}\right)\frac{\piipii}{m_2^2}
\nonumber\\[1ex]&\quad
+\left(\frac{698723}{19200}+\frac{21745 \pi ^2}{16384}\right) \frac{\npi^2}{m_1^2}
+\left(\frac{63641 \pi^2}{24576}-\frac{2712013}{19200}\right) \frac{\npi \npii}{m_1 m_2}
\nonumber\\[1ex]&\quad
+\left(\frac{3200179}{57600}-\frac{28691 \pi ^2}{24576}\right)\frac{\npii^2}{m_2^2},
\\[3ex]
H_{40}(\mathbf{x}_a,\mathbf{p}_a) &=
-\frac{m_1^4}{16}
+\left(\frac{6237 \pi^2}{1024}-\frac{169799}{2400}\right) m_1^3 m_2
+\left(\frac{44825 \pi^2}{6144}-\frac{609427}{7200}\right)m_1^2 m_2^2.
\end{align}
\end{subequations}

The 4PN-accurate dynamics defined by the Hamiltonian \eqref{ah4pn}
[to be augmented by the Galileo-invariant nonlocal piece (\ref{eq5.10})]
is Poincar\'e-invariant in the sense of admitting ten conserved quantities
whose standard Poisson brackets realize the full (PN-expanded) Poincar\'e algebra \cite{Damour:2000kk}.
To prove this, the construction of the (\emph{unique})
boost generator $K^i(\mathbf{x}_a,\mathbf{p}_a,t) = G^i(\mathbf{x}_a,\mathbf{p}_a) - t\,P^i(\mathbf{x}_a,\mathbf{p}_a)$,
with $P^i(\mathbf{x}_a,\mathbf{p}_a) = p_{1i}+p_{2i}$,
and with a center-of-energy vector $G^i(\mathbf{x}_a,\mathbf{p}_a)$,
which can be written as
\be
G^i ({\bf x}_a , {\bf p}_a) = \sum_a \Big( M_a({\bf x}_b , {\bf p}_b) \, x_a^i 
+ N_a({\bf x}_b , {\bf p}_b) \, p_{ai} \Big),
\ee
is crucial.
The functions $M_a$ and $N_a$ possess the following 4PN-accurate expansions
\begin{subequations}
\begin{align}
M_a &= m_a + \frac{1}{c^2}\,M_a^{\rm 1PN} 
+ \frac{1}{c^4} \, M_a^{\rm 2PN} + \frac{1}{c^6} \, M_a^{\rm 3PN} + \frac{1}{c^8} \, M_a^{\rm 4PN},
\\[1ex]
N_a &= \frac{1}{c^4} \, N_a^{\rm 2PN} + \frac{1}{c^6} \, N_a^{\rm 3PN} + \frac{1}{c^8} \, N_a^{\rm 4PN}.
\end{align}
\end{subequations}
The 3PN-accurate parts of these expansions were constructed in Ref.\ \cite{Damour:2000kk}.
For completeness we give here their explicit expressions. They read
\begin{subequations}
\begin{align}
M_1^{\rm 1PN}  &=
\frac{1}{2}\frac{\pipi}{m_1}
- \frac{1}{2}\frac{Gm_1m_2}{r_{12}},
\\[2ex]
M^{\text{2PN}}_1 &= -\frac{1}{8}\frac{\pipip^2}{m_1^3}
+ \frac{1}{4}\frac{Gm_1m_2}{r_{12}} \bigg( -5\,\frac{\pipi}{m_1^2}
- \frac{\pjpj}{m_2^2} + 7\,\frac{\pipj}{m_1m_2} + \frac{\npi\npj}{m_1m_2} 
\bigg)
\nonumber\\[2ex]&\quad
+ \frac{1}{4}\frac{Gm_1m_2}{r_{12}}\frac{G(m_1+m_2)}{r_{12}},
\\[2ex]
M^{\text{3PN}}_1 &=
\frac{1}{16}\frac{\pipip^3}{m_1^5}
+ \frac{1}{16} \frac{Gm_1m_2}{r_{12}} \Bigg(
9\,\frac{\pipip^2}{m_1^4}
+ \frac{\pjpjp^2}{m_2^4}
- 11\,\frac{\pipi\,\pjpj}{m_1^2m_2^2}
- 2\,\frac{\pipj^2}{m_1^2m_2^2}
+ 3\,\frac{\pipi\,\npj^2}{m_1^2m_2^2}
\nonumber\\[2ex]&\quad
+ 7\,\frac{\pjpj\,\npi^2}{m_1^2m_2^2}
- 12\,\frac{\pipj\,\npi\npj}{m_1^2m_2^2}
- 3\,\frac{\npi^2\npj^2}{m_1^2m_2^2} \Bigg)
\nonumber\\[2ex]&\quad
+ \frac{1}{24}\frac{G^2m_1m_2}{r_{12}^2} \Bigg(
(112m_1+45m_2)\frac{\pipi}{m_1^2}
+ (15m_1+2m_2)\frac{\pjpj}{m_2^2}
- \frac{1}{2}(209m_1+115m_2)\frac{\pipj}{m_1m_2}
\nonumber\\[2ex]&\quad
- (31m_1+5m_2)\frac{\npi\npj}{m_1m_2}
+ \frac{\npi^2}{m_1}
- \frac{\npj^2}{m_2}
\Bigg)
\nonumber\\[2ex]&\quad
- \frac{1}{8} \frac{Gm_1m_2}{r_{12}}\frac{G^2 (m_1^2+5m_1m_2+m_2^2)}{r_{12}^2},
\end{align}
\end{subequations}
and
\begin{subequations}
\begin{align}
N^{\text{2PN}}_1 &= -\frac{5}{4}\, G\, \npj,
\\[2ex]
N^{\text{3PN}}_1 &= \frac{1}{8}\frac{G}{m_1m_2}
\Big( 2\,\pipj\npj - \pjpj\,\npi + 3\,\npi\npj^2 \Big)
\nonumber\\[2ex]&\quad
+ \frac{1}{48}\frac{G^2}{r_{12}}
\Big( 19\,m_2\,\npi + \left(130\,m_1+137\,m_2\right)\npj \Big).
\end{align}
\end{subequations}

We have extended the method of undetermined coefficients employed at the 3PN level in Ref.\ \cite{Damour:2000kk}
to the next 4PN level and have found  unique functions $M_a^{\rm 4PN}$ and $N_a^{\rm 4PN}$.
The function $M_1^{\rm 4PN}$ has the structure
\begin{align}
M_1^{\rm 4PN}(\mathbf{x}_a,\mathbf{p}_a) &= -\frac{5\pipip^4}{128 m_1^7}
+ \frac{G m_1 m_2}{r_{12}} M_{46}(\mathbf{x}_a,\mathbf{p}_a)
+ \frac{G^2 m_1 m_2}{r_{12}^2} \Big(m_1\,M_{441}(\mathbf{x}_a,\mathbf{p}_a) + m_2\,M_{442}(\mathbf{x}_a,\mathbf{p}_a)\Big)
\nonumber\\[1ex]&\quad
+ \frac{G^3 m_1 m_2}{r_{12}^3} \Big( m_1^2\,M_{421}(\mathbf{x}_a,\mathbf{p}_a)
+ m_1 m_2\,M_{422}(\mathbf{x}_a,\mathbf{p}_a) + m_2^2\,M_{423}(\mathbf{x}_a,\mathbf{p}_a) \Big)
\nonumber\\[1ex]&\quad
+ \frac{G^4 m_1 m_2}{r_{12}^4}M_{40}(\mathbf{x}_a,\mathbf{p}_a),
\end{align}
where
\begin{subequations}
\begin{align}
M_{46}(\mathbf{x}_a,\mathbf{p}_a) &=
-\frac{13 \pipip^3}{32 m_1^6}
-\frac{15 \npi^4 \npii^2}{256 m_1^4m_2^2}
+\frac{45 \npi^2 \npii^2 \pipi}{128 m_1^4m_2^2}
-\frac{91 \npii^2 \pipip^2}{256 m_1^4 m_2^2}
\nonumber\\[1ex]&\quad
-\frac{5\npi^3 \npii \pipii}{32 m_1^4 m_2^2}
+\frac{25 \npi\npii \pipi \pipii}{32 m_1^4 m_2^2}
+\frac{5 \npi^2\pipii^2}{64 m_1^4 m_2^2}
\nonumber\\[1ex]&\quad
+\frac{7 \pipi \pipii^2}{64 m_1^4m_2^2}
+\frac{11 \npi^4 \piipii}{256 m_1^4 m_2^2}
-\frac{47\npi^2 \pipi \piipii}{128 m_1^4 m_2^2}
+\frac{91 \pipip^2\piipii}{256 m_1^4 m_2^2}
\nonumber\\[1ex]&\quad
+\frac{5 \npi^3 \npii^3}{32 m_1^3m_2^3}
-\frac{7 \npi \npii^3 \pipi}{32 m_1^3 m_2^3}
+\frac{15 \npi^2 \npii^2 \pipii}{32 m_1^3 m_2^3}
\nonumber\\[1ex]&\quad
+\frac{7 \npii^2\pipi \pipii}{32 m_1^3 m_2^3}
-\frac{5 \npi \npii\pipii^2}{16 m_1^3 m_2^3}
-\frac{\pipii^3}{16 m_1^3 m_2^3}
-\frac{11\npi^3 \npii \piipii}{32 m_1^3 m_2^3}
\nonumber\\[1ex]&\quad
+\frac{7 \npi\npii \pipi \piipii}{32 m_1^3 m_2^3}
-\frac{5 \npi^2 \pipii\piipii}{32 m_1^3 m_2^3}
+\frac{\pipi \pipii \piipii}{32 m_1^3m_2^3}
\nonumber\\[1ex]&\quad
+\frac{15 \npi^2 \npii^4}{256 m_1^2 m_2^4}
-\frac{11\npii^4 \pipi}{256 m_1^2 m_2^4}
+\frac{5 \npi \npii^3\pipii}{32 m_1^2 m_2^4}
-\frac{5 \npii^2 \pipii^2}{64 m_1^2m_2^4}
\nonumber\\[1ex]&\quad
-\frac{21 \npi^2 \npii^2 \piipii}{128 m_1^2 m_2^4}
+\frac{7\npii^2 \pipi \piipii}{128 m_1^2 m_2^4}
-\frac{\npi \npii\pipii \piipii}{32 m_1^2 m_2^4}
\nonumber\\[1ex]&\quad
+\frac{\pipii^2 \piipii}{64 m_1^2m_2^4}
+\frac{11 \npi^2 \piipiip^2}{256 m_1^2 m_2^4}
+\frac{37 \pipi\piipiip^2}{256 m_1^2 m_2^4}
-\frac{\piipiip^3}{32 m_2^6},
\\[2ex]
M_{441}(\mathbf{x}_a,\mathbf{p}_a) &=
\frac{7711 \npi^4}{3840 m_1^4}
-\frac{2689 \npi^2 \pipi}{3840m_1^4}
+\frac{2683 \pipip^2}{1920 m_1^4}
-\frac{67 \npi^3 \npii}{30m_1^3 m_2}
\nonumber\\[1ex]&\quad
+\frac{1621 \npi \npii \pipi}{1920 m_1^3m_2}
-\frac{411 \npi^2 \pipii}{1280 m_1^3 m_2}
-\frac{25021 \pipi\pipii}{3840 m_1^3 m_2}
\nonumber\\[1ex]&\quad
+\frac{289 \npi^2 \npii^2}{128 m_1^2m_2^2}
-\frac{259 \npii^2 \pipi}{128 m_1^2 m_2^2}
+\frac{689\npi \npii \pipii}{192 m_1^2 m_2^2}
+\frac{11 \pipii^2}{48m_1^2 m_2^2}
\nonumber\\[1ex]&\quad
-\frac{147 \npi^2 \piipii}{64 m_1^2 m_2^2}
+\frac{283\pipi \piipii}{64 m_1^2 m_2^2}
+\frac{7 \npi \npii^3}{12m_1 m_2^3}
+\frac{49 \npii^2 \pipii}{48 m_1 m_2^3}
\nonumber\\[1ex]&\quad
-\frac{7\npi \npii \piipii}{6 m_1 m_2^3}
-\frac{7 \pipii \piipii}{48m_1 m_2^3}
-\frac{9 \piipiip^2}{32 m_2^4},
\\[2ex]
M_{442}(\mathbf{x}_a,\mathbf{p}_a) &=
-\frac{45 \pipip^2}{32 m_1^4}
+\frac{7 \pipi \pipii}{48 m_1^3m_2}
+\frac{7 \npi \npii \pipi}{6 m_1^3 m_2}
-\frac{49\npi^2 \pipii}{48 m_1^3 m_2}
\nonumber\\[1ex]&\quad
-\frac{7 \npi^3 \npii}{12 m_1^3 m_2}
+\frac{7 \pipii^2}{24 m_1^2 m_2^2}
+\frac{635 \pipi\piipii}{192 m_1^2 m_2^2}
-\frac{983 \npi^2 \piipii}{384 m_1^2m_2^2}
\nonumber\\[1ex]&\quad
+\frac{413 \npi^2 \npii^2}{384 m_1^2 m_2^2}
-\frac{331\npii^2 \pipi}{192 m_1^2 m_2^2}
+\frac{437 \npi \npii\pipii}{64 m_1^2 m_2^2}
\nonumber\\[1ex]&\quad
+\frac{11 \npi \npii^3}{15 m_1m_2^3}
-\frac{1349 \npii^2 \pipii}{1280 m_1 m_2^3}
-\frac{5221\npi \npii \piipii}{1920 m_1 m_2^3}
\nonumber\\[1ex]&\quad
-\frac{2579 \pipii\piipii}{3840 m_1 m_2^3}
+\frac{6769 \npii^2 \piipii}{3840m_2^4}
-\frac{2563 \piipiip^2}{1920 m_2^4}
-\frac{2037 \npii^4}{1280 m_2^4},
\\[2ex]
M_{421}(\mathbf{x}_a,\mathbf{p}_a) &=
-\frac{179843 \pipi}{14400 m_1^2}
+\frac{10223 \pipii}{1200 m_1m_2}
-\frac{15 \piipii}{16 m_2^2}
+\frac{8881 \npi \npii}{2400 m_1m_2}
+\frac{17737 \npi^2}{1600 m_1^2},
\\[2ex]
M_{422}(\mathbf{x}_a,\mathbf{p}_a) &=
\left(\frac{8225 \pi ^2}{16384}-\frac{12007}{1152}\right)\frac{\pipi}{m_1^2}
+\left(\frac{143}{16}-\frac{\pi ^2}{64}\right)\frac{\pipii}{m_1m_2}
+\left(\frac{655}{1152}-\frac{7969 \pi ^2}{16384}\right)\frac{\piipii}{m_2^2}
\nonumber\\[1ex]&\quad
+\left(\frac{6963 \pi ^2}{16384}-\frac{40697}{3840}\right)\frac{\npi^2}{m_1^2}
+\left(\frac{119}{16}+\frac{3 \pi ^2}{64}\right) \frac{\npi\npii}{m_1 m_2}
\nonumber\\[1ex]&\quad
+\left(\frac{30377}{3840}-\frac{7731 \pi ^2}{16384}\right)\frac{\npii^2}{m_2^2},
\\[2ex]
M_{423}(\mathbf{x}_a,\mathbf{p}_a) &=
- \frac{35\pipi}{16 m_1^2}
+ \frac{1327\pipii}{1200 m_1 m_2}
+ \frac{52343\piipii}{14400 m_2^2}
- \frac{2581\npi\npii}{2400 m_1 m_2}
- \frac{15737\npii^2}{1600 m_2^2},
\\[2ex]
M_{40}(\mathbf{x}_a,\mathbf{p}_a) &=
\frac{m_1^3}{16}
+ \left(\frac{3371\pi^2}{6144}-\frac{6701}{1440}\right) m_1^2 m_2
+ \left(\frac{20321}{1440}-\frac{7403\pi^2}{6144}\right) m_1 m_2^2
+ \frac{m_2^3}{16}.
\end{align}
\end{subequations}
The structure of the function $N_1^{\rm 4PN}$ is a bit simpler,
\begin{align}
N_1^{\rm 4PN}(\mathbf{x}_a,\mathbf{p}_a) &= G m_2 N_{45}(\mathbf{x}_a,\mathbf{p}_a)
+ \frac{G^2 m_2}{r_{12}} \Big(m_1\,N_{431}(\mathbf{x}_a,\mathbf{p}_a) + m_2\,N_{432}(\mathbf{x}_a,\mathbf{p}_a)\Big)
\nonumber\\[1ex]&\quad
+ \frac{G^3 m_2}{r_{12}^2} \Big(m_1^2\,N_{411}(\mathbf{x}_a,\mathbf{p}_a)
+ m_1 m_2\,N_{412}(\mathbf{x}_a,\mathbf{p}_a) + m_2^2\,N_{413}(\mathbf{x}_a,\mathbf{p}_a)\Big),
\end{align}
where
\begin{subequations}
\begin{align}
N_{45}(\mathbf{x}_a,\mathbf{p}_a) &=
-\frac{5 \npi^3 \npii^2}{64 m_1^3 m_2^2}
+\frac{\npi \npii^2\pipi}{64 m_1^3 m_2^2}
+\frac{5 \npi^2 \npii \pipii}{32m_1^3 m_2^2}
\nonumber\\[1ex]&\quad
-\frac{\npii \pipi \pipii}{32 m_1^3m_2^2}
+\frac{3 \npi \pipii^2}{32 m_1^3 m_2^2}
-\frac{\npi^3\piipii}{64 m_1^3 m_2^2}
-\frac{\npi \pipi \piipii}{64 m_1^3m_2^2}
\nonumber\\[1ex]&\quad
+\frac{\npi^2 \npii^3}{32 m_1^2 m_2^3}
-\frac{7 \npii^3\pipi}{32 m_1^2 m_2^3}
+\frac{3 \npi \npii^2 \pipii}{16m_1^2 m_2^3}
+\frac{\npii \pipii^2}{16 m_1^2 m_2^3}
\nonumber\\[1ex]&\quad
-\frac{9\npi^2 \npii \piipii}{32 m_1^2 m_2^3}
+\frac{5 \npii \pipi\piipii}{32 m_1^2 m_2^3}
-\frac{3 \npi \pipii \piipii}{16 m_1^2m_2^3}
-\frac{11 \npi \npii^4}{128 m_1 m_2^4}
\nonumber\\[1ex]&\quad
+\frac{\npii^3\pipii}{32 m_1 m_2^4}
+\frac{7 \npi \npii^2 \piipii}{64 m_1m_2^4}
+\frac{\npii \pipii \piipii}{32 m_1 m_2^4}
-\frac{3\npi \piipiip^2}{128 m_1 m_2^4},
\\[2ex]
N_{431}(\mathbf{x}_a,\mathbf{p}_a) &=
-\frac{387 \npi^3}{1280 m_1^3}
+\frac{10429 \npi \pipi}{3840m_1^3}
-\frac{751 \npi^2 \npii}{480 m_1^2 m_2}
+\frac{2209\npii \pipi}{640 m_1^2 m_2}
\nonumber\\[1ex]&\quad
-\frac{6851 \npi \pipii}{1920m_1^2 m_2}
+\frac{43 \npi \npii^2}{192 m_1 m_2^2}
-\frac{125\npii \pipii}{192 m_1 m_2^2}
+\frac{25 \npi \piipii}{48 m_1m_2^2}
\nonumber\\[1ex]&\quad
-\frac{7 \npii^3}{8 m_2^3}
+\frac{7 \npii \piipii}{12 m_2^3},
\\[2ex]
N_{432}(\mathbf{x}_a,\mathbf{p}_a) &=
\frac{7 \npii \pipi}{48 m_1^2 m_2}
+\frac{7 \npi \pipii}{24m_1^2 m_2}
-\frac{49 \npi^2 \npii}{48 m_1^2 m_2}
+\frac{295\npi \npii^2}{384 m_1 m_2^2}
\nonumber\\[1ex]&\quad
-\frac{5 \npii \pipii}{24m_1 m_2^2}
-\frac{155 \npi \piipii}{384 m_1 m_2^2}
-\frac{5999\npii^3}{3840 m_2^3}
+\frac{11251 \npii \piipii}{3840 m_2^3},
\\[2ex]
N_{411}(\mathbf{x}_a,\mathbf{p}_a) &=
-\frac{37397 \npi}{7200 m_1}-\frac{12311 \npii}{2400 m_2},
\\[2ex]
N_{412}(\mathbf{x}_a,\mathbf{p}_a) &=
\left(\frac{5005 \pi ^2}{8192}-\frac{81643}{11520}\right)\frac{\npi}{m_1}
+\left(\frac{773 \pi ^2}{2048}-\frac{61177}{11520}\right)\frac{\npii}{m_2},
\\[2ex]
N_{413}(\mathbf{x}_a,\mathbf{p}_a) &=
-\frac{7073 \npii}{1200 m_2}.
\end{align}
\end{subequations}

\end{widetext}

\end{document}